\documentclass[structabstract]{aa}  
\usepackage{graphicx}
\usepackage{array}
\usepackage{txfonts}
\usepackage{natbib}
\bibpunct{(}{)}{;}{a}{}{,}

\begin{document}
   \title{Multiwavelength campaign on Mrk 509}
   \subtitle{I. Variability and spectral energy distribution}

\author{J.S. Kaastra\inst{1,2}
  \and P.-O. Petrucci\inst{3}
  \and M. Cappi\inst{4}
  \and N. Arav\inst{5}
  \and E. Behar\inst{6}
  \and S. Bianchi\inst{7}
  \and J. Bloom\inst{8}
  \and A.J. Blustin\inst{9}
  \and G. Branduardi-Raymont\inst{10}
  \and E. Costantini\inst{1}
  \and M. Dadina\inst{4}
  \and R.G. Detmers\inst{1,2}
  \and J. Ebrero\inst{1}
  \and P.G. Jonker\inst{1,11,12}
  \and C. Klein\inst{8}
  \and G.A. Kriss\inst{13,14}
  \and P. Lubi\'nski\inst{15}
  \and J. Malzac\inst{16,17}
  \and M. Mehdipour\inst{10}
  \and S. Paltani\inst{18}
  \and C. Pinto\inst{1}
  \and G. Ponti\inst{19}
  \and E.M. Ratti\inst{1}
  \and R.A.N. Smith\inst{10}
  \and K.C. Steenbrugge\inst{20,21}
  \and C.P. de Vries\inst{1}
  }
  
\institute{SRON Netherlands Institute for Space Research, Sorbonnelaan 2,
           3584 CA Utrecht, the Netherlands 
	   \and
	   Sterrenkundig Instituut, Universiteit Utrecht, 
	   P.O. Box 80000, 3508 TA Utrecht, the Netherlands
	   \and
	   UJF-Grenoble 1 / CNRS-INSU, Institut de Plan\'etologie et d'Astrophysique
	   de Grenoble (IPAG) UMR 5274, Grenoble, F-38041, France
           \and
	   INAF-IASF Bologna, Via Gobetti 101, 40129 Bologna, Italy
	   \and
	   Department of Physics, Virginia Tech, Blacksburg, VA 24061, USA
	   \and
	   Department of Physics, Technion-Israel Institute of Technology, 
	   Haifa 32000, Israel 	
	   \and 
	   Dipartimento di Fisica, Universit\`a degli Studi Roma Tre, 
	   via della Vasca Navale 84, 00146 Roma, Italy 
	   \and
	   Department of Astronomy, 601 Campbell Hall, 
	   University of California, Berkeley, CA 94720, USA
	   \and
	   Institute of Astronomy, University of Cambridge, 
	   Madingley Road, Cambridge CB3 0HA, UK
	   \and
	   Mullard Space Science Laboratory, University College London, 
	   Holmbury St. Mary, Dorking, Surrey, RH5 6NT, UK
	   \and
	   Harvard-Smithsonian Center for Astrophysics, 60 Garden Street, 
	   Cambridge, MA 02138, USA
	   \and
	   Department of Astrophysics, IMAPP, Radboud University Nijmegen, 
	   PO Box 9010, NL-6500 GL Nijmegen, the Netherlands
	   \and
	   Space Telescope Science Institute, 3700 San Martin Drive, 
	   Baltimore, MD 21218, USA
	   \and
	   Department of Physics and Astronomy, The Johns Hopkins University,
	   Baltimore, MD 21218, USA
	   \and
	   Centrum Astronomiczne im. M. Kopernika, 
	   Rabia\'nska 8, PL-87-100 Toru\'n, Poland
	   \and
	   Universit\'e de Toulouse, UPS-OMP, IRAP, Toulouse, France
	   \and
           CNRS, IRAP, 9 Av. colonel Roche, BP 44346, F-31028 Toulouse 
	   cedex 4, France
	   \and
	   ISDC Data Centre for Astrophysics, Astronomical Observatory of the
	   University of Geneva, 16, ch. d'Ecogia, 1290 Versoix, Switzerland 
	   \and
	   School of Physics and Astronomy, University of Southampton, 
	   Highfield, Southampton SO17 1BJ
	   \and
           Instituto de Astronom\'ia, Universidad Cat\'olica del Norte, 
	   Avenida Angamos 0610, Casilla 1280, Antofagasta, Chile
	   \and
	   Department of Physics, University of Oxford, Keble Road, 
	   Oxford OX1 3RH, UK
} 
\date{\today}

\abstract
{Active galactic nuclei show a wealth of interesting physical processes, some of
which are poorly understood. In a broader context, they play an important role
in processes that are far beyond their immediate surroundings, owing to the high
emitted power.}
{We want to address a number of open questions, including the location and
physics of the outflow from AGN, the nature of the continuum emission, the
geometry and physical state of the X-ray broad emission line region, the Fe-K
line complex, the metal abundances of the nucleus and finally the interstellar
medium of our own Galaxy as seen through the signatures it imprints on the 
X-ray and UV spectra of AGN.}
{We study one of the best targets for these aims, the Seyfert 1 galaxy Mrk~509
with a multiwavelength campaign using five satellites (XMM-Newton, INTEGRAL,
Chandra, HST and Swift) and two ground-based facilities (WHT and PAIRITEL). Our
observations cover more than five decades in frequency, from 2~$\mu$m to
200~keV. The combination of high-resolution spectroscopy and time variability
allows us to disentangle and study the different components. Our campaign covers
100 days from September to December 2009, and is centred on a simultaneous
set of deep XMM-Newton and INTEGRAL observations with regular time intervals,
spanning seven weeks.}
{We obtain a continuous light curve in the X-ray and UV band, showing a strong,
up to 60\% flux increase in the soft X-ray band during the three weeks in the
middle of our deepest monitoring campaign, and which is correlated with an
enhancement of the UV flux. This allows us to study the time evolution of the
continuum and the outflow. By stacking the observations, we have also obtained
one of the best X-ray and UV spectra of a Seyfert galaxy ever obtained. In this
paper we also study the effects of the spectral energy distribution (SED) that
we  obtained on the photo-ionisation equilibrium. Thanks to our broad-band
coverage, uncertainties on the SED do not strongly affect the determination of
this equilibrium.}
{Here we present our very successful campaign and in a series of subsequent
papers we will elaborate on different aspects of our study.}

\keywords{Galaxies: active --  quasars: absorption lines -- X-rays: general}
\maketitle

\section{Introduction}

Active galactic nuclei (AGN) are compact sources with very high luminosities,
located at the centres of galaxies. Accretion onto the super-massive black holes
(SMBH) at their centres is generally believed to be the driving process for the
activity. Thanks to their brightness, they form one of the richest laboratories
for studying astrophysical processes. In this paper we present one of the
deepest multiwavelength campaigns of an AGN, the Seyfert 1 galaxy
\object{Mrk~509}. It is among the best for these studies because it is unique in
combining X-ray brightness, outflow features, and significant but moderate
variability. Below we introduce the most important astrophysical processes that
are addressed by our study. 

After our introduction of the relevant astrophysics, we briefly provide an
overview in Sect.~\ref{sect:mrk509} of the target of our campaign, Mrk~509,
followed by an overview of the observations in Sect.~\ref{sect:log}, and we
present the spectral energy distribution (SED) and light curve in the subsequent
sections. Details about various aspects of our study are deferred to subsequent
papers of this series. 

\section{Astrophysics of AGN in the context of our campaign}

\subsection{Outflows from AGN}

In recent years the potential importance of AGN outflows for the growth of
super-massive black holes \citep{silk1998,blandford1999,king2003,blandford2004},
the enrichment of the intergalactic medium \citep{furlanetto2001,cavaliere2002},
the evolution of the host galaxy \citep{scannapieco2004}, cluster cooling flows
\citep{wu2000,bower2001,ciotti2001,borgani2002,platania2002}, the magnetisation
of cluster and galactic gas \citep{daly1990,furlanetto2001,kronberg2001}, and
the luminosity function of AGN \citep{wyithe2003} has been widely recognised.
However, for the lack of a better alternative, theoretical studies use the
physical properties of the outflow (metallicity, mass, and kinetic energy flux)
as free parameters because there are few observational constraints. To assess
the importance of AGN outflows on the processes mentioned above, it is essential
to establish the real mass flux, metallicity, and kinetic luminosity
($\dot{E}_{\mathrm k}$) of AGN winds. Determining $\dot{E}_{\mathrm k}$ requires
answers to some fundamental questions. What is the physical state of the
outflowing gas and what is its total column density? What is the distance of the
gas from the central source?

The $\dot{E}_{\mathrm k}$ of a shell-like, non-accelerating outflow is given by
\begin{equation}
\dot{E}_{\mathrm k}\simeq {1\over 2} \Omega f_m R N_{\mathrm H} m_{\mathrm p} v^3,
\end{equation}
where $\Omega$ is the solid angle occupied by the outflow, $R$ the distance from
the central source, $N_{\mathrm H}$ the total hydrogen column density,
$m_{\mathrm p}$ the proton mass, $f_m=\rho/n_{\rm H}m_{\rm p}$ (1.43 for a
plasma with proto-solar abundances) with $\rho$ the mass density and $n_{\rm H}$
the hydrogen density, and $v$ the outflow velocity. Spectral observations
straightforwardly determine $v$, and $\Omega$ is expected to be $\sim \pi$ since
50\% of all Seyfert 1s show outflow signatures \citep{crenshaw1999}. Our
campaign focuses on determining the two most uncertain quantities: $N_{\mathrm
H}$ and $R$. Determining these is tightly connected to the physical state and
location of the outflow, as explained below.

In the best studied AGN at least two to three ionisation components are needed
to model the rich X-ray absorption spectrum. In \object{NGC~3783} there are
three discrete components, which are most likely in pressure equilibrium,
representing different phases of gas at the same distance from the nucleus
\citep{krongold2003,netzer2003}. However, in \object{NGC~5548} at least five
ionisation components are needed if the X-ray absorber is modelled by a finite
number of discrete absorption systems, and these absorbers cannot be in pressure
equilibrium \citep{kaastra2002}. Instead, a continuous, power-law distribution
of $N_{\mathrm H}$ versus ionisation parameter $\xi$ gives a better description
of the data \citep{steenbrugge2003,steenbrugge2005}. In yet another case,
Mrk~279, the distribution is continuous but more complicated than a simple
power-law \citep{costantini2007}. 

Photo-ionisation modelling of the outflow yields the ionisation parameter
$\xi\equiv L/nR^2$, where $L$ is the 1--1000~Ryd ionising luminosity and $n$ the
hydrogen density. Independent measurements of $n$ then determine $R$. The use of
density sensitive X-ray lines is difficult and has not (yet) delivered robust
results \citep{kaastra2004}. Alternatively, when $L$ changes, the outflow has to
adjust to the new situation. How fast this happens depends on the recombination
time scale, which scales as $\sim n^{-1}$. This method has been applied to a
280~ks XMM-Newton observation of NGC~3783 with wildly different results. From
the RGS data $R>$10~pc was inferred based on the lack of change in the deep
\ion{O}{vii}/\ion{O}{viii} absorption edges and the Fe-M UTA absorption complex
\citep{behar2003}. On the other hand, variability in the \ion{Fe}{xxv} resonance
line at 6.7~keV detected in the EPIC data \citep{reeves2004} implies $R<$0.2~pc.
From a 100~ks XMM-Newton observation of \object{NGC~4051}, $R\simeq 0.001$~pc
was deduced for the highest ionisation gas \citep{krongold2007}. However, the
large amplitude variations ($\max/\min=12$) and the short time scales (down to
100~s) make these results rather model-dependent; an analysis of Chandra LETGS
data yielded distances in the range of 0.02--1~pc \citep{steenbrugge2009}.

Density-sensitive lines have given more robust results in the UV, and in at
least one case (NGC~3783) they yield a distance comparable to the one determined
from densities based on recombination time scales. Using metastable transitions
in [\ion{C}{iii}] $\lambda$1176, \citet{gabel2005} find a distance of
$\sim$25~pc for the absorbjing gas in one of the components in NGC~3783. They
find a comparable distance based on the recombination time for \ion{Si}{iv}.
Again using the [\ion{C}{iii}] metastable transitions, gas in one of the
components in \object{NGC~4151} lies at a distance of $<$0.1~pc
\citep{kraemer2006}. Using metastable levels in [\ion{Si}{ii} and
[\ion{Fe}{ii}], \citet{moe2009} find a distance for a low-ionisation outflow in
the quasar \object{SDSS~J0838+2955} of $\sim 3$~kpc. Similarly, large distances
for outflows have been found using similar density diagnostics in luminous
quasars \citep{hamann2001,dunn2010,aoki2011,arav2011}.

In order to obtain a reliable density, we need both accurate ionic column
densities, which yield the ionisation structure and total outflowing column
density, and time variability on a suitable time scale, from which the location
of the outflow can be constrained through monitoring. With the current
generation of X-ray telescopes, only a handful of AGN with outflows have high
enough fluxes to yield suitable data for such an ambitious programme. Of these
targets, Mrk~509 is the most promising due to the following three attributes. 1)
It is among the brightest X-ray AGNs in the sky. 2) It has an excellent X-ray
line structure \citep{pounds2001,smith2007,detmers2010}, not as deep and blended
as NGC~3783, but not as shallow and sparse as Mrk~279. Mrk~509 is a
representative example of a moderate outflow. 3) Mrk~509 varies on a time scale
of a few days making it ideal for a spectral timing campaign. Other targets like
\object{MCG~$-6$-30-15}, \object{Mrk~766}, or NGC~4051 with long XMM-Newton
exposure times vary too rapidly to do this: within the characteristic time
scales of 100--1000~s, it is simply not possible to get a high-quality grating
spectrum.

\subsection{Kinematics of the outflow as traced in the UV}

Our campaign on Mrk~509 includes HST high-resolution ultraviolet spectral
observations that are simultaneously with Chandra's grating X-ray observations.
The ultraviolet spectra provide a more detailed view of the kinematics of the
outflowing gas. Prior high-resolution spectral observations with the Far
Ultraviolet Spectroscopic Explorer (FUSE) \citep{kriss2000} and the Space
Telescope Imaging Spectrograph (STIS) on the Hubble Space Telescope (HST)
\citep{kraemer2003} provide a good baseline for our observations, but these were
both obtained nearly a decade before to our current campaign. The new UV spectra
obtained with the Cosmic Origins Spectrograph (COS) on HST provide an updated
view of the outflow, and enable us to examine long-term characteristics of the
variability.

Although the UV absorbers are not necessarily a perfect tracer of the X-ray
absorbing gas \citep{kriss2000,kraemer2003}, they provide insight into both
lower ionisation absorbing components and portions of the high-ionisation
outflow. Our COS observations yield accurate measurements of (or limits on) the
column densities of \ion{C}{ii}, \ion{C}{iv}, \ion{N}{v}, \ion{O}{i},
\ion{Si}{ii} -- \ion{Si}{iv}, \ion{S}{ii}, \ion{S}{iii}, \ion{Fe}{ii},
\ion{Fe}{iii}, and \ion{H}{i} Ly$\alpha$. Our limits or detections on the
low-ionisation ions of Fe, Si, and S nicely complement the high-ionisation range
covered by the RGS and the LETGS. Velocity-resolved measurements of the Li-like
doublets of \ion{C}{iv} and \ion{N}{v} can be used to determine column densities
and covering fractions of the outflowing gas \citep[e.g.,][]{arav2007}. 

\subsection{Abundances in AGN}

Studying the abundances, especially those of C, N, O, and Fe, of the gas in
galaxies informs us about the ongoing enrichment processes through AGB star
winds and supernova type Ia and II explosions. Therefore there has long been an
interest in studying abundances with redshift to determine the star formation
rate through history. Obvious sources for the study of the abundances are AGN,
by taking advantage of their luminosity. Generally, the abundances in AGN have
been determined from broad emission lines, however several assumptions make this
method subject to systematic errors. Thanks to the improvement in atomic data,
especially in the dielectronic recombination rates for iron, the excellent
statistics in the combined RGS spectrum, and the simultaneous optical and hard
X-ray fluxes to constrain the spectral energy distribution (SED), we can
determine reliable and accurate relative abundances from the narrow absorption
lines of the outflow observed in Mrk~509.

Since absolute abundances (i.e., metal abundances relative to hydrogen) cannot
be measured directly from X-ray spectra, our UV observations of the Ly$\alpha$
line provide a constraint on the total hydrogen column density. The first
absolute abundance estimate for an outflow was derived by our group from the UV
spectra of Mrk~279, where absorption lines of H, C, N, and O were carefully
modelled \citep{arav2007} to find abundances relative to solar for carbon
(2.2$\pm$0.7), nitrogen, (3.5$\pm$1.1) and oxygen (1.6$\pm$0.8), which fully
agree with the relative abundances of C, N, and O derived from the simultaneous
X-ray spectra \citep{costantini2007}. The apparently enhanced N/H and N/O ratios
in the Mrk~279 outflow may hint at strong contributions from stellar winds from
massive stars and AGB stars.

\subsection{Broad emission lines}

Broad lines, especially at the energy of the \ion{O}{vii} triplet and the
\ion{O}{viii} Ly$\alpha$ line, have been recently detected  in high-resolution
X-ray spectra
\citep{kaastra2002,ogle2004,steenbrugge2005,costantini2007,longinotti2010}. The
width of the lines (about 1~\AA, 10\,000~km\,s$^{-1}$) is comparable to the
width of the broad lines detected in the UV band and known to be produced in the
so-called broad line region. In at least one case, the soft X-ray emission lines
could be related, via a physical model \citep[the locally optimally emitting
clouds model,][]{baldwin1995}, to the UV lines \citep{costantini2007}. The
contribution of the broad line region gas to the iron K$\alpha$ line at 6.4~keV
is not completely understood. For classical Seyfert~1 galaxies this contribution
seems modest \citep[e.g. $<$20\% in Mrk~279,][]{costantini2010}, while it may
account for most of the line emission in  specific objects \citep[e.g.
\object{NGC~7213},][]{bianchi2008}. The quality of the data of Mrk~509 allows us
to significantly detect broad emission features around the \ion{O}{vii} resonant
line, the \ion{O}{viii}, \ion{N}{vii} Ly$\alpha$ lines and the \ion{Ne}{ix}
triplet. The simultaneous observation of HST/COS and Chandra/LETGS allow us to
connect the broad lines detected in the UV with the soft energy lines and even
the iron K$\alpha$ line.

\subsection{Iron K complex}

Mrk~509 also shows a rich variety of emission and absorption components in the
Fe-K band. In particular, EPIC data from previous XMM-Newton observations shows
evidence of a complex Fe-K emission line, with a narrow and neutral component
possibly produced far from the source, plus a broad, ionised, and variable
component possibly originating in the accretion disk
\citep{pounds2001,page2003,ponti2009}. 

In addition, strong absorption features were found in the same data set at
rest-frame energies 8--8.5~keV and 9.7~keV. These were interpreted as being
produced by H-like iron K and K-shell absorptions associated with an outflow
with mildly relativistic velocity of $0.14-0.2 c$. The lines were found to be
variable in energy and marginal in intensity, implying that variations in
either the column density, geometry, and/or ionisation structure of the outflow
were maybe common in this source \citep{dadina2005,cappi2009}.

The above properties, combined with the source brightness
($F_{2-10\,{\mathrm{keV}}} \sim 2-5\times 10^{-14}$~W\,m$^{-2}$) make Mrk~509
unique in attempting time-resolved spectral studies in the Fe-K band to follow
the time evolution of both emission and absorption Fe-K features, and in
disentangling the different physical components present in this source.

\subsection{Continuum emission of Seyfert galaxies}

The physical process at the origin of the X-ray emission of Seyfert galaxies is
generally believed to be thermal Comptonisation. In this process, the soft UV
photons coming from the accretion disk are up-scattered into the X-ray domain by
the hot thermal electrons of the corona. A thermal distribution is generally
preferred to a non-thermal one since the discovery, more than 15 years ago, of a
high-energy cut-off near 100 keV in NGC~4151 by OSSE and Sigma
\citep{jourdain1992,maisack1993}. Such a cut-off was not expected by the
non-thermal models developed at that time to explain the power-law-like X-ray 
spectrum \citep[e.g.][]{zdziarski1990}. The presence of a high-energy cut-off
has now been observed in a large number of objects
\citep{perola2002,beckmann2005,dadina2008}.

The cold (disk) and hot (corona) phases are expected to be radiatively linked:
part of the cold emission, which gives birth to the UV bump, is produced by the
reprocessing of part of the hot emission. Inversely the hot emission, at the
origin of the broad-band X-ray continuum, is believed to be produced by Compton
up-scattering of the soft photons, emitted by the cold phase, on the coronal
energetic electrons. The system must then satisfy equilibrium energy balance
equations, which depend on geometry and on the ratio of direct heating of the
disk to that of the corona \citep{haardt1991}. In the limiting case of a
`passive' disk, the amplification of the Comptonisation process is only fixed by
geometry. Therefore, if the corona is in energy balance, its temperature $T_{\rm
e}$ and optical depth $\tau$ must satisfy a relation that can be computed for
different geometries of the disk $+$ corona configuration
\citep[e.g.,][]{svensson1996}. 

Realistic thermal Comptonisation spectra have been computed for more than two
decades, and important effects like the anisotropy of the soft photon field have
been precisely taken into account \citep{haardt1993,stern1995,poutanen1996}.
These effects appear far from negligible. Noticeably, the spectral shape is
significantly  different for different disk-corona geometries but also for 
different viewing angles. Moreover, these effects underline the differences
between realistic thermal Comptonisation spectra and the cut-off power-law
approximation generally used to mimic them \citep{petrucci2001}. This may have
important consequences in the study of superimposed spectral components like the
soft X-ray excess and/or the the iron line and the reflection hump, which
require a precise determination of the underlying continuum.

A well known characteristic of thermal Comptonisation spectra, however, is that
they are strongly degenerate, i.e., significantly different combinations of
temperature and optical depth of the corona give the same power law slope in the
2--10~keV band. To break this degeneracy requires broad-band observations from
UV to hard X-rays, the X-ray/$\gamma$-ray shape constraining the high-energy
cut-off, which is directly linked to the corona temperature. The ratio of the UV
to soft X-ray flux constrains the optical depth better. Stronger and less
ambiguous constraints on the nature of the coronal plasma can also be obtained
from multiwavelength (from UV to hard X-rays/$\gamma$-rays) variability studies,
as they provide direct insight into the way the emitting particles are heated
and cooled. Indeed, variations in the X-ray spectral shape may be produced by
intrinsic changes in the hot corona properties (e.g. changes of the heating
process efficiency) and/or by variations in the external environment, such as
changes in the soft photon flux (and consequently in the coronal cooling)
produced by the cold phase \citep{malzac2000,petrucci2000}. For example, thermal
Comptonisation models, where hot and cold phases are in radiative equilibrium,
predict that the X-ray spectrum of the sources should harden when the energy of
the high-energy cut-off increases \citep[e.g.][]{haardt1997}. Indeed a
correlation between $\Gamma$ and $T_{\rm e}$ has been observed in different
objects \citep{petrucci2000,zdziarski2001}, thus giving strong support to a
thermal nature of the corona particle distribution. The analysis of the one
month, simultaneous IUE/RXTE monitoring campaign on \object{NGC~7469}, performed
in 1996, is also in agreement with thermal Comptonisation emission
\citep{nandra2000,petrucci2004}.

The XMM-Newton/INTEGRAL monitoring of Mrk~509 presented here provides an ideal
data set to test Comptonisation models and to derive further constraints on the
physical parameters and geometry of the source, as well as on the precise shape
of spectral components such as the soft-X-ray excess, the outflow, and the
reflection component. It should also be noted that a good knowledge of the SED
up to 100 keV is very important input for the photo-ionisation modelling of the
absorption/emission features produced by these outflows.

\section{Mrk~509\label{sect:mrk509}}

Mrk~509 was detected as a Seyfert 1 galaxy with a photographic magnitude
$m_{\mathrm{pg}}=13$ and a size of 10\arcsec\ \citep{1973Afz.....9..487M}. It
has a redshift of 0.034397 \citep{huchra1993} and broad (FWHM
4\,500~km\,s$^{-1}$) and narrow optical emission lines
\citep{1973Afz.....9....5M}. Its optical luminosity is high for a Seyfert 1
galaxy, and it puts Mrk~509 close to the limit between Seyfert galaxies and QSOs
\citep{1974Afz....10..483K}.

The host galaxy extends approximately east-west \citep{1976Afz....12..431M} and
has an axial ratio $b/a\sim 0.85$
\citep{1988ApJS...67..249D,1990AJ.....99.1435K}. Mrk~509 presently shows no
nearby neighbours \citep{1976Afz....12..431M}, the closest spiral galaxy is at a
projected distance of 0.3~Mpc \citep{1988AJ.....96.1235F},  but the asymmetric
halo to the south \citep{1990ApJS...72..231M} is direct evidence for strong
interactions at some point in the recent past.

In almost any wavelength band that opened its window, Mrk~509 has been one of
the first AGN to be studied, thanks to its brightness, and several well-known
astrophysicists today have investigated this object in earlier stages of their
careers.

In the radio band, Mrk~509 was discovered at 3.9~cm in 1976
\citep{1978Afz....14...91M}. Spatially resolved measurements at 6 and 20~cm show
an asymmetric structure of 1.4\arcsec $\times $0.5\arcsec\ in PA 124\degr\
\citep{1984ApJ...278..544U,1987MNRAS.228..521U}, corresponding to a linear size
of about 1~kpc. This core is surrounded by more diffuse emission, visible at
20~cm, with a size comparable to the optical size ($10\times 8$\arcsec, or
$7\times 5$~kpc \citep{1992A&A...264..489S}. The radio emission seems to be
preferentially aligned at PA $-40$\degr\ to $-60$\degr, coinciding with the
preferred direction of the optical polarisation
\citep{1992A&A...264..489S,1983ApJ...266..470M}.

In the infrared the first observations date back to 1975
\citep{1976ApJ...205...44S,1976ApJ...207..367A}. The spectrum in the
1--10\,$\mu$m band is approximately a power law
\citep{1978ApJ...226..550R,1982A&A...107..276G,1983ApJS...52..341M}. Contrary to
several other Seyferts, Mrk~509 shows no evidence of strong dust features in
infrared spectra \citep{1984MNRAS.207...35R,1986A&A...166....4M}. The
variability in the mid-IR is much weaker than at shorter wavelengths, with a
possible delay of two to three months \citep{2004MNRAS.350.1049G}. This suggests
that the bulk of the IR emission originates far from the nucleus. 

Quantitative measurements of the optical broad and narrow emission lines  were
first presented by \citet{1977ApJ...215..733O} and \citet{1978AJ.....83.1257D}.
Since then, dozens of papers have appeared with refined measurements and models
for the line profiles and intensity ratios. The size of the broad emission line
region in the H$\beta$ line has been estimated as 80 light days (0.07~pc) based
on reverberation studies \citep{1998ApJ...501...82P}.

Perhaps the most detailed study of interest for our project has been the mapping
of the [\ion{O}{iii}] $\lambda 5007$ line by \citet{1983ApJ...274..558P}. They
show that the line has two components. There is a rotating, low-ionisation gas
disk, coinciding with hot stellar components, which is photo-ionised by the UV
radiation of young hot stars. The rotation axis has approximately the same
direction as the radio emission and the preferred polarisation angle of the
optical emission mentioned before. The second component is expanding,
high-ionisation gas, which is photo-ionised by the nucleus. The line profile of
that component is consistent with outflowing gas up to velocities of
$-800$~km\,s$^{-1}$. It extends out to about 5~kpc from the nucleus.  Outflowing
gas is seen at opposite sides of the nucleus, indicating that the outflow is
approximately face-on. A similar spatial extent is seen in a few other narrow
emission lines \citep{2000MNRAS.316....1W}. Approximately the same region also
shows broad H$\beta$ line emission, most likely caused by scattered nuclear
light in the narrow line region \citep{1998ApJ...494L...9M}. X-rays from the
nucleus travel through these media, and the lowewst ionised X-ray absorbers may
show the imprint of this material on the spectrum.

The first UV spectra were presented in 1980, with measurements of the Ly$\alpha$
line flux using the IUE satellite \citep{1980ApJ...242...14W}. Significantly
better spectra were subsequently obtained with HST
\citep{1995AJ....110.1026C,crenshaw1999,kraemer2003} and FUSE \citep{kriss2000}.

In X-rays, the first detection was made with Ariel~V in 1974--1976 in the 2--10
keV band \citep{1978MNRAS.182..489C}. Soon the first (power-law) spectra were
fitted \citep{1980ApJ...235..377M}, and the source was detected at higher
energies, beyond 40~keV \citep{1981ApJ...250..513D} and up to about 200~keV
\citep{1981A&A....94..234P,1983ApJ...269..423R}. It was one of the first sources
where a soft X-ray excess was discovered with HEAO1-A2
\citep{1985ApJ...297..633S}, the first iron line detection dated to 1987 with
EXOSAT \citep{1987ApJ...317..145M}, and the reflection component was revealed by
Ginga \citep{1994MNRAS.267..193P}.

Variability was discovered first in the optical band
\citep{1976Afz....12..431M}. They found variations of about a magnitude on a
time scale of several months, but a comparison with observations 21 years before
showed a similar flux level. This is one of the desired properties for the
purpose of our campaign: significant variability on suitable time scales, with a
limited range of flux levels. Since then, variability has been found in all
energy bands from radio \citep{1978Afz....14...91M}, infrared continuum
\citep{2004MNRAS.350.1049G}, optical polarisation \citep{1983ApJ...266..470M},
broad emission lines
\citep{1982ApJS...49..469P,1984ApJ...279..529P,1992ApJS...81...59R}, UV
\citep{1985ApJ...297..151C}, and X-rays \citep{1980ApJ...235..355D}.

There have been many papers that derive the mass of the central SMBH using a
variety of methods. Derived numbers range from $1.4\times 10^9$~M$_{\odot}$ in
one of the oldest determinations, based on the correlation of the H$\beta$
intensity with the full-width at zero intensity \citep{1983AcApS...3..113L} to
$1.43\pm 0.12\times 10^8$~M$_{\odot}$ for a modern estimate based on
reverberation mapping of optical broad lines \citep{2004ApJ...613..682P}.

\section{Observations\label{sect:log}}

\begin{table*}[!htbp]
\begin{minipage}[t]{\hsize}
\setlength{\extrarowheight}{3pt}
\caption{The observation log of our Mrk 509 campaign.}
\label{tab:obslog}
\centering
\renewcommand{\footnoterule}{}
\tiny
\begin{tabular}{c c c c c || c c c c c}
\hline \hline
 & & & \multicolumn{1}{c}{Start time (UTC)} & Span &  & & & \multicolumn{1}{c}{Start time (UTC)} & Span  \\
Telescope & Obs. & ID & yyyy-mm-dd@hh:mm & (ks) & Telescope & Obs. & ID & yyyy-mm-dd@hh:mm & (ks) \\ 
\hline

Chandra & 1 & 11387 & 2009-12-10@04:54 & 131.4				& HST/COS & 1 & lbdh01010 & 2009-12-10@02:48 & 2.0 \\                               
Chandra & 2 & 11388 & 2009-12-12@22:33 & 48.6 				& HST/COS & 2 & lbdh01020 & 2009-12-10@04:07 & 2.7 \\	
\cline{1-5}
XMM-Newton & 1 & 0601390201 & 2009-10-15@06:19 & 57.1 			& HST/COS & 3 & lbdh01030 & 2009-12-10@05:42 & 5.5 \\
XMM-Newton & 2 & 0601390301 & 2009-10-19@15:20 & 59.9 			& HST/COS & 4 & lbdh01040 & 2009-12-10@08:54 & 2.7 \\	
XMM-Newton & 3 & 0601390401 & 2009-10-23@05:41 & 60.5 			& HST/COS & 5 & lbdh02010 & 2009-12-11@02:46 & 2.0 \\
XMM-Newton & 4 & 0601390501 & 2009-10-29@06:55 & 60.5 			& HST/COS & 6 & lbdh02020 & 2009-12-11@04:05 & 2.7 \\
XMM-Newton & 5 & 0601390601 & 2009-11-02@02:46 & 62.4 			& HST/COS & 7 & lbdh02030 & 2009-12-11@05:41 & 2.7 \\
XMM-Newton & 6 & 0601390701 & 2009-11-06@07:00 & 62.7 			& HST/COS & 8 & lbdh02040 & 2009-12-11@07:16 & 5.5 \\ \cline{6-10}
XMM-Newton & 7 & 0601390801 & 2009-11-10@08:42 & 60.5			& WHT & 1 & 1372032--154 & 2009-10-04@22:50 & 1.6 \\
XMM-Newton & 8 & 0601390901 & 2009-11-14@08:27 & 60.5 			& WHT & 2 & 1372732--166 & 2009-10-07@20:10 & 3.6 \\
XMM-Newton & 9 & 0601391001 & 2009-11-18@02:08 & 65.1			& WHT & 3 & 1378102--127 & 2009-11-01@20:55 & 2.4 \\
XMM-Newton & 10 & 0601391101 & 2009-11-20@07:40 & 62.4 			& WHT & 4 & 1378849--900 & 2009-11-06@19:30 & 5.4 \\
\cline{1-5}
INTEGRAL &  1 & 07200160001 & 2009-10-14@21:55 & 127.7 			& WHT & 5 & 1382682--702 & 2009-11-27@19:42 & 1.5 \\ \cline{6-10}
INTEGRAL &  2 & 07200160002 & 2009-10-19@19:38 & 120.1 			& Swift & 1 & 00035469005 & 2009-09-04@13:18 & 0.9 \\
INTEGRAL &  3 & 07200160003 & 2009-10-22@19:29 & 124.3 			& Swift & 2 & 00035469006 & 2009-09-08@02:30 & 1.5 \\
INTEGRAL &  4 & 07200160004 & 2009-10-28@19:08 & 110.7 			& Swift & 3 & 00035469007 & 2009-09-12@18:52 & 0.5 \\
INTEGRAL &  5 & 07200160005 & 2009-11-01@01:07 & 122.3 			& Swift & 4 & 00035469008 & 2009-09-16@20:36 & 1.1 \\
INTEGRAL &  6 & 07200160006 & 2009-11-05@15:00 & 63.0 			& Swift & 5 & 00035469009 & 2009-09-20@07:51 & 1.1 \\
INTEGRAL &  7 & 07200160011 & 2009-11-06@18:32 & 59.0 			& Swift & 6 & 00035469010 & 2009-09-24@09:51 & 1.3 \\
INTEGRAL &  8 & 07200160007 & 2009-11-09@18:18 & 120.7 			& Swift & 7 & 00035469011 & 2009-10-02@18:45 & 1.0 \\
INTEGRAL &  9 & 07200160008 & 2009-11-13@20:13 & 122.3 			& Swift & 8 & 00035469012 & 2009-10-05@23:59 & 1.0 \\
INTEGRAL & 10 & 07200160009 & 2009-11-17@14:48 & 61.2			& Swift & 9 & 00035469013 & 2009-10-10@21:37 & 1.1 \\
INTEGRAL & 11 & 07200160010 & 2009-11-19@11:23 & 110.7 			& Swift & 10 & 00035469014 & 2009-10-14@10:09 & 1.4 \\
\cline{1-5}
PAIRITEL & 1 & 166.2 & 2009-09-15@04:45 & 0.8						& Swift & 11 & 00035469015 & 2009-10-18@05:56 & 1.0 \\
PAIRITEL & 2 & 166.2 & 2009-09-16@04:35 & 0.8 						& Swift & 12 & 00035469016 & 2009-11-20@07:18 & 1.0 \\
PAIRITEL & 3 & 166.3 & 2009-09-20@04:51 & 0.8 						& Swift & 13 & 00035469017 & 2009-11-24@23:39 & 1.0 \\
PAIRITEL & 4 & 166.4 & 2009-09-24@03:55 & 0.8 						& Swift & 14 & 00035469018 & 2009-11-28@19:13 & 1.2 \\
PAIRITEL & 5 & 166.5 & 2009-09-28@03:24 & 0.8 						& Swift & 15 & 00035469019 & 2009-12-02@00:18 & 1.3 \\
PAIRITEL & 6 & 166.7 & 2009-10-22@02:55 & 0.8 						& Swift & 16 & 00035469020 & 2009-12-06@07:07 & 1.2 \\
PAIRITEL & 7 & 166.8 & 2009-10-27@04:06 & 0.8 						& Swift & 17 & 00035469021 & 2009-12-08@02:30 & 1.1 \\
PAIRITEL & 8 & 166.9 & 2009-10-31@03:21 & 0.8 						& Swift & 18 & 00035469022 & 2009-12-10@12:36 & 1.2 \\
PAIRITEL & 9 & 166.10 & 2009-11-17@03:19 & 0.8 						& Swift & 19 & 00035469023 & 2009-12-12@20:30 & 1.0 \\
\hline
\end{tabular}
\end{minipage}
\end{table*}

We have obtained data from seven different observatories during our campaign. At
the core of our programme are ten observations of approximately 60~ks each, with
XMM-Newton, spaced by four days. We used the data of all instruments on
XMM-Newton: Reflection Grating Spectrometer (RGS), EPIC (pn and MOS), and
Optical Monitor (OM). For this last instrument, all filters except for the white
light filter were used, and we also obtained a spectrum with the optical grism
for each observation. Observations with the UV grism were not allowed for
operational reasons. 

Simultaneously with our ten XMM-Newton observations, we obtained data with
INTEGRAL to observe the hard X-rays. Three weeks after the end of the XMM-Newton
monitoring, we did simultaneous observations with Chandra (Low Energy
Transmission Grating (LETGS) with the HRC-S camera), together with the Hubble
Space Telescope (HST) Cosmic Origins Spectrograph (COS). Due to operational
constraints, these observations could not coincide with the XMM-Newton/INTEGRAL
monitoring.

Before our XMM-Newton/INTEGRAL observations, and in between the XMM-Newton and
Chandra/HST observations, we monitored Mrk~509 with Swift, using both its X-ray
telescope (XRT) and the UltraViolet and Optical Telescope (UVOT). This was in
order to obtain a continuous monitoring and to allow us to study the flux before
the start of our campaign. This is important because absorption components may
respond with a delay to continuum variations.

\begin{figure}
\resizebox{\hsize}{!}{\includegraphics[angle=0]{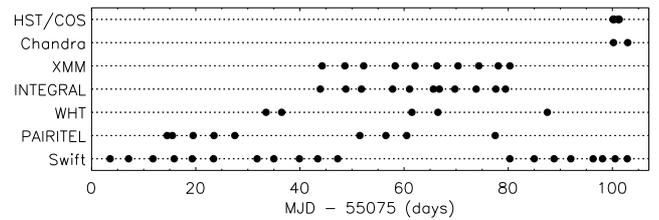}}
\caption{Timeline of our monitoring campaign of Mrk~509. The first observation,
with Swift, started on September 4, 2009, and the last observation, with Chandra,
ended on December 13, 2009.}
\label{fig:timeline}
\end{figure}

Finally, throughout our high-energy campaign, we obtained optical spectroscopy
and photometry. We have five observations with the 4.2~m William Herschel
Telescope (WHT) at La Palma, photometry with the ACAM camera using the Sloan g,
r, i and Z filters, and low-resolution spectroscopy with the VPH disperser.
Furthermore, we obtained nine observations with the 1.3~m PAIRITEL (Peters
Automated IR Imaging Telescope), with photometry in the J, H, and K bands.
Table~\ref{tab:obslog} gives some details on our observations, and in
Fig.~\ref{fig:timeline} we show a graphical overview of the timeline of our
campaign.

\section{Light curve}

\begin{figure}
\resizebox{\hsize}{!}{\includegraphics[angle=0]{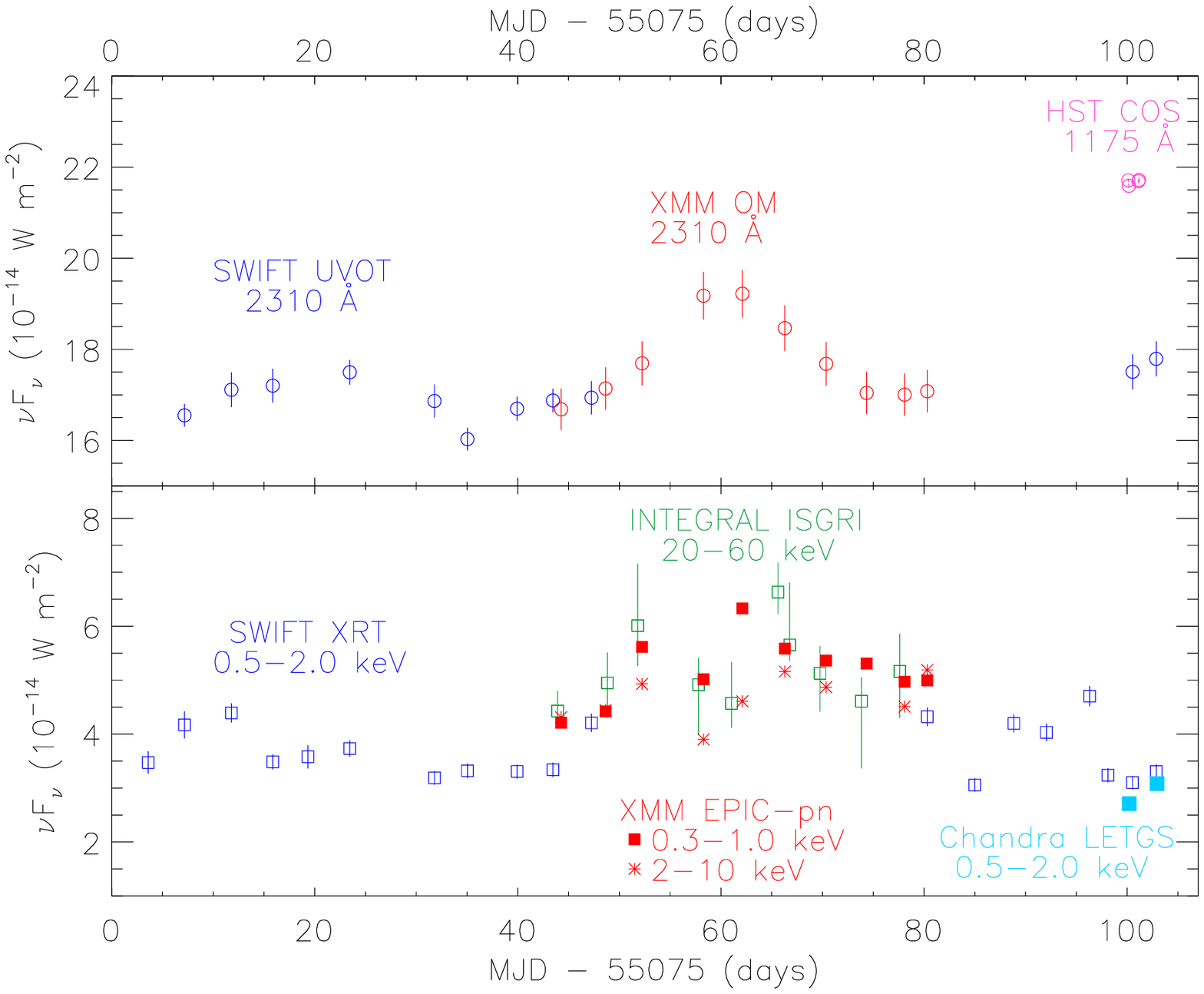}}
\caption{Top panel: UV light curve of Mrk~509 as obtained with Swift/UVOT,
XMM-Newton/OM, and HST/COS. The data have been corrected for extinction. Bottom
panel: X-ray light curve as obtained by Swift/XRT, XMM-Newton/pn, Chandra/LETGS,
and INTEGRAL/ISGRI. These data have also been corrected for Galactic absorption.
For the pn data, the squares correspond to the 0.3--1.0 keV band, the stars to
the 2--10 keV band. Note that $10^{12}$~JyHz equals $10^{-14}$~W\,m$^{-2}$
($10^{-11}$ in c.g.s. units).}
\label{fig:lightcurve}
\end{figure}

The light curves in the X-ray band (0.3--1.0~keV, 2--10~keV and 20--60~keV) and
in UV (2310~\AA) are shown in Fig.~\ref{fig:lightcurve}. More details and the
interpretation of these light curves are given in subsequent papers of our
series \citep[e.g.][]{mehdipour2011,petrucci2011}. The flux was at a typical
flux level, compared with archival observations. Right in the middle of our
observing campaign, the source showed a significant brightening by $\sim 50$\%,
in particular in the soft X-ray band, and with a smaller amplitude in the UV and
hard X-ray bands. This allowed us to study the response of the absorption
components to this brightening, and thanks to our broad-band measurements we can
constrain the emission mechanisms for the continuum components. 

\section{Spectral energy distribution\label{section:sed}}

The broad-band spectral energy distribution (SED) is essential for obtaining the
ionisation balance needed for the photo-ionisation modelling of the outflow. The
multiwavelength nature of our campaign gave us almost simultaneous coverage of
the total spectrum, which allowed us to constrain the exact shape of the SED to
high accuracy. 

\begin{figure}[htbp]
\resizebox{\hsize}{!}{\includegraphics[angle=-90]{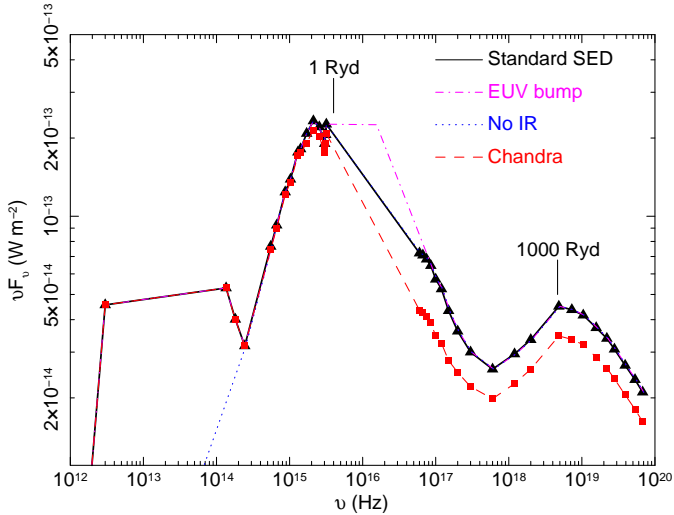}}
\caption{\label{fig:sed}
The SED used of Mrk~509 for the time-averaged XMM-Newton and Chandra
observations. Data points from Table~\ref{tab:sed} are indicated with triangles
and squares.}
\end{figure}

We need two different SEDs for our campaign. The first describes the average SED
during the \textit{XMM-Newton} part of the campaign, which is needed for the
analysis of the stacked RGS spectrum. The second one is needed for the Chandra
LETGS and HST COS part of the campaign, when the source had a lower flux
level. Table~\ref{tab:sed} shows the flux points of every instrument used to
create the SED shown in Fig.~\ref{fig:sed}. The statistical uncertainties on the
flux are much lower than the reported values (except in the case of INTEGRAL),
therefore we list the systematic uncertainties due to calibration
uncertainties. 

\begin{table}
\caption{Continuum fluxes corrected for Galactic and intrinsic absorption
during the XMM-Newton and Chandra parts of the campaign.} 
\label{tab:sed}
\centering                        
\begin{tabular}{llcccl} 
\hline\hline
Frequency & $E/\lambda$ & $\nu F_\nu^{\ \ \ a}$ & $\nu F_\nu^{\ \ \ b}$ & Unc.$^{c}$ &
Instr.$^{d}$ \\
(Hz)      &       &               &                  &  (\%)       & \\
\hline
3.91$\times 10^{19}$ &  162~keV    & 2.67 &  2.05  & 50 &INTEGRAL \\
2.79$\times 10^{19}$ &  115~keV    & 3.08 &  2.37  & 30 &INTEGRAL \\
2.20$\times 10^{19}$ &   91~keV    & 3.39 &  2.60  & 10 &INTEGRAL \\
1.57$\times 10^{19}$ &   65~keV    & 3.72 &  2.86  & 10 &INTEGRAL \\
1.05$\times 10^{19}$ &   43~keV    & 4.16 &  3.20  & 10 &INTEGRAL \\
7.30$\times 10^{18}$ &   30~keV    & 4.37 &  3.36  & 10 &INTEGRAL \\
4.84$\times 10^{18}$ &   20~keV    & 4.50 &  3.46  & 10 &INTEGRAL \\
2.00$\times 10^{18}$ &  1.5~\AA\   & 3.35 &  2.58  &  5 &pn \\
1.20$\times 10^{18}$ &  2.5~\AA\   & 2.95 &  2.27  &  5 &pn \\
6.00$\times 10^{17}$ &  5.0~\AA\   & 2.58 &  1.98  &  5 &pn/RGS \\ 
3.00$\times 10^{17}$ &  10~\AA\    & 3.00 &  2.21  &  5 &RGS \\
2.00$\times 10^{17}$ &  15~\AA\    & 3.61 &  2.51  &  5 &RGS \\ 
1.50$\times 10^{17}$ &  20~\AA\    & 4.33 &  2.78  &  5 &RGS \\ 
1.20$\times 10^{17}$ &  25~\AA\    & 5.25 &  3.24  &  5 &RGS \\ 
1.00$\times 10^{17}$ &  30~\AA\    & 5.72 &  3.46  &  5 &RGS \\ 
8.57$\times 10^{16}$ &  35~\AA\    & 6.46 &  3.89  &  5 &RGS \\ 
7.50$\times 10^{16}$ &  40~\AA\    & 6.83 &  4.12  &  5 &LETGS \\ 
6.67$\times 10^{16}$ &  45~\AA\    & 7.08 &  4.27  &  5 &LETGS \\ 
6.00$\times 10^{16}$ &  50~\AA\    & 7.21 &  4.35  &  5 &LETGS \\ 
3.18$\times 10^{15}$ &  943~\AA\   & 22.6 &  20.8  & 10 &FUSE \\ 
3.13$\times 10^{15}$ &  960~\AA\   & 20.6 &  19.0  & 10 &FUSE \\ 
3.05$\times 10^{15}$ &  984~\AA\   & 20.6 &  18.8  & 10 &FUSE \\ 
3.02$\times 10^{15}$ &  993~\AA\   & 19.0 &  17.5  & 10 &FUSE \\ 
2.55$\times 10^{15}$ &  1175~\AA\  & 22.1 &  20.3  &  5 &FUSE \\ 
2.12$\times 10^{15}$ &  1415~\AA\  & 23.3 &  21.5  &  5 &COS \\ 
1.71$\times 10^{15}$ &  1750~\AA\  & 20.8 &  19.1  &  5 &COS \\
1.42$\times 10^{15}$ &  2120~\AA\  & 18.2 &  17.5  &  3 &OM \\
1.30$\times 10^{15}$ &  2310~\AA\  & 17.6 &  17.1  &  3 &OM \\
1.03$\times 10^{15}$ &  2910~\AA\  & 13.9 &  13.5  &  3 &OM \\
8.72$\times 10^{14}$ &  3440~\AA\  & 12.4 &  12.1  &  3 &OM \\
6.67$\times 10^{14}$ &  4500~\AA\  & 9.24 &  8.98  &  3 &OM \\
5.52$\times 10^{14}$ &  5430~\AA\  & 7.65 &  7.46  &  4 &OM \\
2.44$\times 10^{14}$ &  1.23~$\mu$m& \multicolumn{2}{c}{3.17}  & 10 &J$^{e}$\\
1.81$\times 10^{14}$ &  1.66~$\mu$m& \multicolumn{2}{c}{4.01}  & 10 &H$^{e}$\\
1.35$\times 10^{14}$ &  2.22~$\mu$m& \multicolumn{2}{c}{5.30}  & 10 &K$^{e}$\\
3.00$\times 10^{12}$ &  100~$\mu$m & \multicolumn{2}{c}{4.56}  & 15 &IRAS$^{e}$ \\
3.00$\times 10^{10}$ &  1~cm       &\multicolumn{2}{c}{4.56$\times 10^{-7}$} & 15 & Radio$^{e}$ \\
\hline                          
\end{tabular}
\smallskip
\begin{list}{}{}
\item[$^{\mathrm{a}}$] For the time-averaged XMM-Newton spectrum, in units of
$10^{-14}$~W\,m$^{-2}$ (or 5.034~photons\,m$^{-2}$\,s$^{-1}$\,\AA$^{-1}$)
\item[$^{\mathrm{b}}$] For the time-averaged Chandra spectrum, in units of
$10^{-14}$~W\,m$^{-2}$ (or 5.034~photons\,m$^{-2}$\,s$^{-1}$\,\AA$^{-1}$)
\item[$^{\mathrm{c}}$] Estimated systematic and statistical uncertainty
\item[$^{\mathrm{d}}$] Instrument used to derive the flux 
\item[$^{\mathrm{e}}$] Values obtained from the literature for the J, H, K,
IRAS and radio bands; see text for references.
\end{list}
\end{table}                        
                                   
\subsection{X-ray SED}             
                                   
We used the best fit RGS model \citep[model 2 of][]{detmers2011} for the
spectrum between 7--38~\AA. We correct for both the intrinsic absorption (the
ionised outflow) and the Galactic ISM absorption. For the flux between 1.2~\AA\
(10 keV) and 7~\AA\ (1.77~keV), we used the EPIC-pn data. Above 10~keV we used
the INTEGRAL data, which were obtained simultaneous with the XMM-Newton
observations. This gives us the continuum flux up to $\sim 200$~keV. Beyond
that, the SED is extrapolated using the model described in
\citet{petrucci2011}. 

We compared the Chandra LETGS flux in two different energy bands with the
non-contemporaneous RGS flux to obtain the flux variations in the soft
($20-35$~\AA) and hard ($7-10$~\AA) X-ray bands. The RGS flux in these bands is
66\% and 30\% higher than the LETGS flux, respectively. The soft X-ray flux for
the RGS observation for $\lambda>37$~\AA\ is obtained by scaling the LETGS
continuum by a factor of 1.66. As we have no information on the X-ray flux above
10~keV during the LETGS observation, and the uncertainty on the LETGS spectrum
for $\lambda<7$~\AA\ increases, we estimate the flux for the Chandra observation
for $\lambda<7$~\AA\ by dividing our model for XMM-Newton/INTEGRAL by a factor
of 1.30. 

\subsection{EUV extrapolation\label{sect:euvsed}}

The EUV spectrum produces most of the ionising flux; however, it is also the
most uncertain part of the SED, as there are no data between 50 and 912~\AA.
LETGS formally measures up to 175~\AA, but due to the Galactic absorption, the
flux is low and also the modelling of the higher spectral orders becomes more
uncertain at longer wavelengths. We therefore have to interpolate our SED
between the soft X-rays and the UV. There are several options for doing this.
Our baseline model is a power-law interpolation between 50--943~\AA\ (the last
FUSE data point). For the XMM-Newton epoch this power law has a photon index of
2.39, for the Chandra epoch 2.53. Alternatively, we consider some kind of big
blue bump, by extrapolating the RGS spectrum with a power law, using its slope
near 30~\AA\ (photon index 2.58) up to the point (193~\AA) where it matches the
extrapolation from the UV with a photon index of $-2$. The different
extrapolations between the X-ray and UV data are shown in Fig.~\ref{fig:sed}.

\subsection{UV SED}

The UV part of the SED was derived from archival FUSE data and from the HST COS
data of our campaign. As FUSE and COS share a common wavelength band, the FUSE
archival fluxes were scaled to the COS flux level, assuming that the spectral
shape remained the same. 

The full description of the COS data reduction can be found in
\citet{kriss2011}. In short, the COS data were flat-field corrected and
additional wavelength calibration performed. Time-dependent sensitivity
corrections were applied to the COS data, resulting in an absolute flux accuracy
of 5\%. As the COS data were taken during the LETGS observations, we needed to
adjust them to the higher flux levels during the XMM-Newton observations. The OM
data at somewhat longer wavelengths (see below) show on average 6\% higher flux
during the XMM-Newton observations compared to the LETGS observations. Using the
near-simultaneous FUSE/HST/optical spectrum of Mrk~509 from \citet{shang2005}
normalised to the OM flux, we find that increasing the COS flux by 10\% is
appropriate if assuming a constant shape for the UV spectrum.

\subsection{Optical SED}

The optical part of the SED was obtained from the OM data. The data were
corrected for interstellar absorption and de-reddened using the reddening curve
of \citet{cardelli1989}, including the near-UV update by \citet{odonnell1994}.
The total colour excess $E(B-V)$ is 0.057~mag, and $R_V=A(V)/E(B-V)$ was fixed
at 3.1. The total neutral hydrogen column density in the direction of Mrk~509 is
$4.44 \times 10^{24}$~m$^{-2}$, as given by \citet{murphy1995}. The host galaxy
correction is based on the results of \citet{bentz2009} and \citet{kinney1996}.
For Mrk~509 the host-galaxy contribution at 2231~\AA\ is $7\times
10^{-20}$~W\,m$^{-2}$\, \AA$^{-1}$, which is negligible compared to the AGN flux
of $8.45\times 10^{-17}$~W\,m$^{-2}$\,\AA$^{-1}$. At longer wavelengths the host
galaxy contribution increases, to $3.2\times 10^{-18}$~W\,m$^{-2}$\,\AA$^{-1}$
at 5500~\AA, where the AGN flux is $1.7\times 10^{-17}$~W\,m$^{-2}$\,\AA$^{-1}$.
For a more thorough description of the optical data reduction, see
\citet{mehdipour2011}. The optical fluxes (from Swift UVOT) at the time of the
LETGS observations have to be adjusted for the lower flux level of the source.
As XMM-Newton observation 2 has a similar flux to that of Swift UVOT observation
18, which is close in time to the LETGS observations, we assume that the optical
flux during the LETGS observations is the same as that of XMM-Newton OM
observation 2, which is 6\% smaller than the average optical flux during the XMM
observations. 

\subsection{Infrared and radio SED}

At IR wavelengths shortward of 1~$\mu$m, the spectrum shows a small upturn due
to emission from the torus \citep[see, e.g., the spectra by][]{landt2011}. The
infrared fluxes that we use here are based on photometry with host-galaxy
subtraction in the J, H, and K bands. We use the average from the observations of
\citet{1992ApJ...399...38D} and \citet{1992MNRAS.256..149K}, which are in good
agreement with each other. For 100~$\mu$m, we use the IRAS flux
\citep{moshir1990}. Between the K-band and the 100~$\mu$m flux points, the SED
shows an almost constant value in terms of $\nu F_{\nu}$. In the 6--35~$\mu$m
band Spitzer data exist \citep{wu2009}, with $\sim 30$\% higher flux compared to
our adopted SED. However, these data, taken with a relatively large aperture of
10\arcsec, are not corrected for the stellar contribution. As our results are
not very sensitive to the details of the IR spectrum, we can safely use our
simplified SED in the mid-IR band as shown in Fig.~\ref{fig:sed}.

Beyond 100~$\mu$m, the flux drops rapidly; following usual practice in this
region, we extrapolate the flux from 100~$\mu$m to lower frequencies with an
energy index of $-2.5$ down to 1~cm. This gives a flux of only about a factor of
two above the observed value at 14.9~GHz \citep{1996AJ....111.1431B}. Finally,
we also consider an SED without IR emission, because it is unclear how much of
the IR emission is seen by the outflow (see next section). 

\subsection{Effect of the SED on the ionisation balance}

\begin{figure}[!htbp]
\resizebox{\hsize}{!}{\includegraphics[angle=-90]{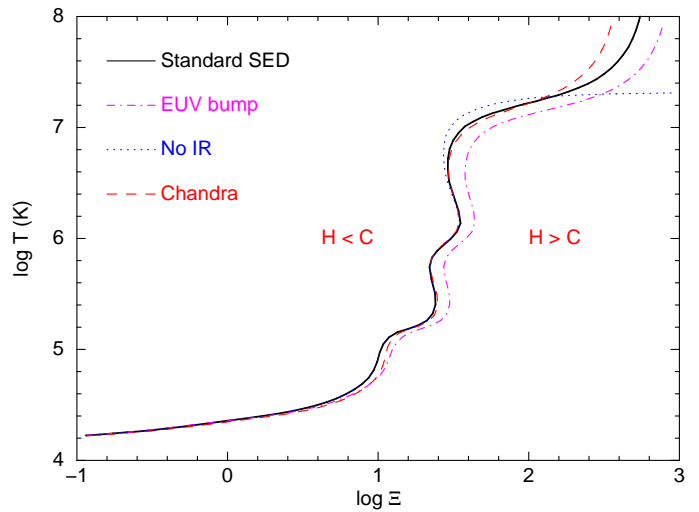}}
\caption{\label{fig:cooling}
The four different cooling curves for the different assumed SEDs.}
\end{figure}

Because a different SED can have a strong effect on the ionisation balance of
the outflow, we investigated the effects of different SEDs, focusing on the
uncertainties in the EUV and infrared parts of the spectrum. In the EUV band we
have no measurements, so we have to rely on interpolation. Depending on the
location of the outflow relative to the dusty torus expected to surround the
AGN, it may receive either all or only a small fraction of the IR flux emitted
by the torus. To investigate the effects of different assumptions about this, we
compare four SEDs (Fig.~\ref{fig:sed}):  

\begin{enumerate}
\item Our standard SED (Table~\ref{tab:sed}, column 3 and Fig.~\ref{tab:sed},
solid line), which uses a simple power-law interpolation between the soft X-ray
and UV bands;
\item A SED with a stronger EUV flux, using the broken power-law approximation
with a break at 193~\AA\ mentioned in Sect.~\ref{sect:euvsed}, to mimic a
stronger EUV flux (EUV bump in Fig.~\ref{fig:sed});
\item Same as model 1, but with the infrared flux essentially set to zero, for
an absorber that does not receive emission from the torus (No IR in
Fig.~\ref{fig:sed});
\item the SED for the Chandra observation (Table~\ref{tab:sed}, column~4), to
see the effects of time variability.
\end{enumerate}

The ionisation balance calculations were performed using 
Cloudy\footnote{http://www.nublado.org/} \citep{ferland1998} version C08.00,
with \citet{lodders2009} abundances. The results are shown in
Fig.~\ref{fig:cooling}. As can be seen, the differences between the four cooling
curves are small. The case without infrared flux (model 3) starts deviating from
the default case above $10^6$~K, because of enhanced cooling by inverse Compton
scattering of the infrared photons. The case with a stronger EUV flux (model 2)
has a similar shape to model 1, but has on average $\log\Xi$ higher by 0.08 due
to the enhanced ionising EUV flux. This results in a small shift towards the
right of the figure. Finally, the lower flux during the Chandra observations
leads to a very similar cooling curve, because the peak of the UV spectrum is
very similar.

We are therefore confident that our derived SED is an accurate description of
the true source continuum of Mrk 509 and that the photo-ionisation results
obtained from the analysis of \citet{detmers2011} are not seriously affected by
the uncertainties in the assumed SED.  

\begin{figure}[htbp]
\includegraphics[angle= -90,width=9cm]{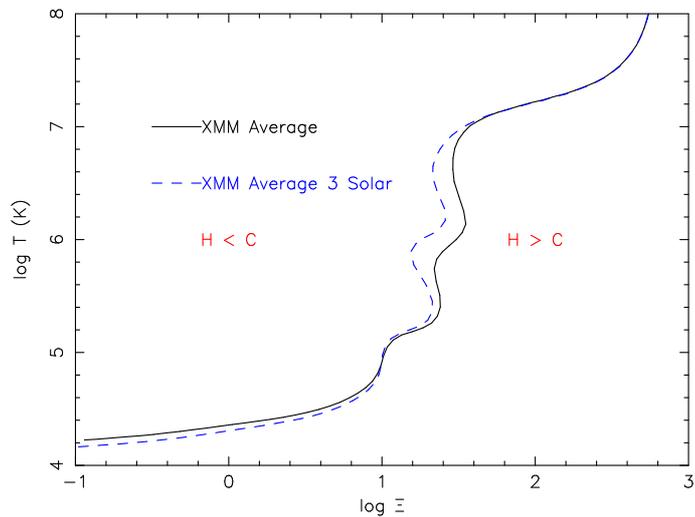}
\caption{\label{fig:metal_cool}
The two different cooling curves for solar (solid line) and 3 times solar
metallicity (dashed line).}
\end{figure} 

As an additional test we checked how metallicity influences the cooling curve.
We did this because we do not know a priori the metallicity of the outflow,
which can have a significant effect on the shape of the cooling curve \citep[see
e.g.][]{chakravorty2009}. We adopted 3 times the solar abundances, cf. typical
abundances found in Mrk~279 \citep{arav2007}. Figure~\ref{fig:metal_cool} shows
the results compared to the solar metallicity run. There are slight differences
in the shape of the two cooling curves, with the higher metallicity curve having
bigger unstable branches (where the slope is negative) due to enhanced line
cooling. However, the resulting photo-ionisation model will give no significant
differences in derived parameters, except for the hydrogen column densities,
which are a factor of three lower owing to the three times higher metal
abundances. This again shows that the ionisation structure of the outflow in
Mrk~509 is very stable to the uncertainties that are present in either the
assumed SED or metallicity.

\section{Summary}

In this paper we have introduced our multiwavelength study of the Seyfert 1
galaxy Mrk~509, which used XMM-Newton, INTEGRAL, Chandra, HST, Swift, WHT and
PAIRITEL. Our observations spanned 100 days of monitoring from September 2009 to
December, covering band-passes from 2~$\mu$m to 200~keV. The core of our
programme consisted of ten simultaneous observations with XMM-Newton and
INTEGRAL, followed by a long Chandra LETGS spectrum obtained simultaneously with
an HST/COS far-UV spectrum. The high-resolution spectroscopy combined with the
time variability monitoring enabled us to disentangle the different absorption
and emission components in Mrk~509. 

Using our data we produced a continuous light curve in the X-ray and UV bands
and a comprehensive SED. Thanks to our broad-band coverage, we deduced that
uncertainties on the SED have little effect on the photo-ionisation equilibrium
that applies to our subsequent models of the ionised outflow. Our light curve
shows a 60\% flux increase in the soft X-ray band correlated with an enhancement
of the UV flux. 

A series of subsequent papers will elaborate on these results and others from
our campaign. The stacked XMM-Newton RGS spectrum is presented in
\citet[][paper~II]{kaastra2011}. The time-averaged ionisation and velocity
structure of the outflow deduced from this spectrum is presented by
\citet[][paper~III]{detmers2011}, the broad-band continuum variability by
\citet[][paper~IV]{mehdipour2011}. \citet[][paper~V]{ebrero2011} describes the
Chandra LETGS data and \citet[][paper~VI]{kriss2011} presents the analysis of
the HST/COS data. More papers are in preparation. 

\begin{acknowledgements}

This work is based on observations obtained with XMM-Newton, an ESA science
mission with instruments and contributions directly funded by ESA Member States
and the USA (NASA). It is also based on observations with INTEGRAL, an ESA
project with instrument and science data centre funded by ESA member states
(especially the PI countries: Denmark, France, Germany, Italy, Switzerland,
Spain), Czech Republic, and Poland and with the participation of Russia and the
USA. This work made use of data supplied by the UK Swift Science Data Centre at
the University if Leicester. SRON is supported financially by NWO, the
Netherlands Organization for Scientific Research. J.S. Kaastra thanks the PI of
Swift, Neil Gehrels, for approving the TOO observations, and the duty scientists
at the William Herschel Telescope for performing the service observations. P.-O.
Petrucci acknowledges financial support from CNES and the French GDR PCHE. M.
Cappi, M. Dadina, S. Bianchi, and G. Ponti acknowledge financial support from
contract ASI-INAF n. I/088/06/0. N. Arav and G. Kriss gratefully acknowledge
support from NASA/XMM-Newton Guest Investigator grant NNX09AR01G. Support for
HST Program number 12022 was provided by NASA through grants from the Space
Telescope Science Institute, which is operated by the Association of
Universities for Research in Astronomy, Inc., under NASA contract NAS5-26555. E.
Behar was supported by a grant from the ISF. A. Blustin acknowledges the support
of a STFC Postdoctoral Fellowship. P. Lubi\'nski has been supported by the
Polish MNiSW grants NN203065933 and 362/1/N-INTEGRAL/2008/09/0. M. Mehdipour
acknowledges the support of a PhD studentship awarded by the UK Science \&
Technology Facilities Council (STFC). G. Ponti acknowledges support via an EU
Marie Curie Intra-European Fellowship under contract no.
FP7-PEOPLE-2009-IEF-254279. K. Steenbrugge acknowledges the support of Comit\'e
Mixto ESO - Gobierno de Chile.

\end{acknowledgements}

\bibliographystyle{aa}
\bibliography{paper1}

\begin{thebibliography}{133}
\expandafter\ifx\csname natexlab\endcsname\relax\def\natexlab#1{#1}\fi

\bibitem[{{Allen}(1976)}]{1976ApJ...207..367A}
{Allen}, D.~A. 1976, \apj, 207, 367

\bibitem[{{Aoki} {et~al.}(2011){Aoki}, {Oyabu}, {Dunn}, {Arav}, {Edmonds},
  {Korista}, {Matsuhara}, \& {Toba}}]{aoki2011}
{Aoki}, K., {Oyabu}, S., {Dunn}, J.~P., {et~al.} 2011, \pasj, in press
  (arXiv1101.4340)

\bibitem[{{Arav} {et~al.}(2011){Arav}, , \& {et~al.}}]{arav2011}
{Arav}, N., , \& {et~al.} 2011, \apj, submitted

\bibitem[{{Arav} {et~al.}(2007){Arav}, {Gabel}, {Korista}, {Kaastra}, {Kriss},
  {Behar}, {Costantini}, {Gaskell}, {Laor}, {Kodituwakku}, {Proga}, {Sako},
  {Scott}, \& {Steenbrugge}}]{arav2007}
{Arav}, N., {Gabel}, J.~R., {Korista}, K.~T., {et~al.} 2007, \apj, 658, 829

\bibitem[{{Baldwin} {et~al.}(1995){Baldwin}, {Ferland}, {Korista}, \&
  {Verner}}]{baldwin1995}
{Baldwin}, J., {Ferland}, G., {Korista}, K., \& {Verner}, D. 1995, \apjl, 455,
  L119+

\bibitem[{{Barvainis} {et~al.}(1996){Barvainis}, {Lonsdale}, \&
  {Antonucci}}]{1996AJ....111.1431B}
{Barvainis}, R., {Lonsdale}, C., \& {Antonucci}, R. 1996, \aj, 111, 1431

\bibitem[{{Beckmann} {et~al.}(2005){Beckmann}, {Shrader}, {Gehrels}, {Soldi},
  {Lubi{\'n}ski}, {Zdziarski}, {Petrucci}, \& {Malzac}}]{beckmann2005}
{Beckmann}, V., {Shrader}, C.~R., {Gehrels}, N., {et~al.} 2005, \apj, 634, 939

\bibitem[{{Behar} {et~al.}(2003){Behar}, {Rasmussen}, {Blustin}, {Sako},
  {Kahn}, {Kaastra}, {Branduardi-Raymont}, \& {Steenbrugge}}]{behar2003}
{Behar}, E., {Rasmussen}, A.~P., {Blustin}, A.~J., {et~al.} 2003, \apj, 598,
  232

\bibitem[{{Bentz} {et~al.}(2009){Bentz}, {Peterson}, {Netzer}, {Pogge}, \&
  {Vestergaard}}]{bentz2009}
{Bentz}, M.~C., {Peterson}, B.~M., {Netzer}, H., {Pogge}, R.~W., \&
  {Vestergaard}, M. 2009, \apj, 697, 160

\bibitem[{{Bianchi} {et~al.}(2008){Bianchi}, {La Franca}, {Matt}, {Guainazzi},
  {Jimenez Bail{\'o}n}, {Longinotti}, {Nicastro}, \&
  {Pentericci}}]{bianchi2008}
{Bianchi}, S., {La Franca}, F., {Matt}, G., {et~al.} 2008, \mnras, 389, L52

\bibitem[{{Blandford} \& {Begelman}(1999)}]{blandford1999}
{Blandford}, R.~D. \& {Begelman}, M.~C. 1999, \mnras, 303, L1

\bibitem[{{Blandford} \& {Begelman}(2004)}]{blandford2004}
{Blandford}, R.~D. \& {Begelman}, M.~C. 2004, \mnras, 349, 68

\bibitem[{{Borgani} {et~al.}(2002){Borgani}, {Governato}, {Wadsley}, {Menci},
  {Tozzi}, {Quinn}, {Stadel}, \& {Lake}}]{borgani2002}
{Borgani}, S., {Governato}, F., {Wadsley}, J., {et~al.} 2002, \mnras, 336, 409

\bibitem[{{Bower} {et~al.}(2001){Bower}, {Benson}, {Lacey}, {Baugh}, {Cole}, \&
  {Frenk}}]{bower2001}
{Bower}, R.~G., {Benson}, A.~J., {Lacey}, C.~G., {et~al.} 2001, \mnras, 325,
  497

\bibitem[{{Cappi} {et~al.}(2009){Cappi}, {Tombesi}, {Bianchi}, {Dadina},
  {Giustini}, {Malaguti}, {Maraschi}, {Palumbo}, {Petrucci}, {Ponti},
  {Vignali}, \& {Yaqoob}}]{cappi2009}
{Cappi}, M., {Tombesi}, F., {Bianchi}, S., {et~al.} 2009, \aap, 504, 401

\bibitem[{{Cardelli} {et~al.}(1989){Cardelli}, {Clayton}, \&
  {Mathis}}]{cardelli1989}
{Cardelli}, J.~A., {Clayton}, G.~C., \& {Mathis}, J.~S. 1989, \apj, 345, 245

\bibitem[{{Cavaliere} {et~al.}(2002){Cavaliere}, {Lapi}, \&
  {Menci}}]{cavaliere2002}
{Cavaliere}, A., {Lapi}, A., \& {Menci}, N. 2002, \apjl, 581, L1

\bibitem[{{Chakravorty} {et~al.}(2009){Chakravorty}, {Kembhavi}, {Elvis}, \&
  {Ferland}}]{chakravorty2009}
{Chakravorty}, S., {Kembhavi}, A.~K., {Elvis}, M., \& {Ferland}, G. 2009,
  \mnras, 393, 83

\bibitem[{{Chapman} {et~al.}(1985){Chapman}, {Geller}, \&
  {Huchra}}]{1985ApJ...297..151C}
{Chapman}, G.~N.~F., {Geller}, M.~J., \& {Huchra}, J.~P. 1985, \apj, 297, 151

\bibitem[{{Ciotti} \& {Ostriker}(2001)}]{ciotti2001}
{Ciotti}, L. \& {Ostriker}, J.~P. 2001, \apj, 551, 131

\bibitem[{{Cooke} {et~al.}(1978){Cooke}, {Ricketts}, {Maccacaro}, {Pye},
  {Elvis}, {Watson}, {Griffiths}, {Pounds}, {McHardy}, {Maccagni}, {Seward},
  {Page}, \& {Turner}}]{1978MNRAS.182..489C}
{Cooke}, B.~A., {Ricketts}, M.~J., {Maccacaro}, T., {et~al.} 1978, \mnras, 182,
  489

\bibitem[{{Costantini} {et~al.}(2007){Costantini}, {Kaastra}, {Arav}, {Kriss},
  {Steenbrugge}, {Gabel}, {Verbunt}, {Behar}, {Gaskell}, {Korista}, {Proga},
  {Quijano}, {Scott}, {Klimek}, \& {Hedrick}}]{costantini2007}
{Costantini}, E., {Kaastra}, J.~S., {Arav}, N., {et~al.} 2007, \aap, 461, 121

\bibitem[{{Costantini} {et~al.}(2010){Costantini}, {Kaastra}, {Korista},
  {Ebrero}, {Arav}, {Kriss}, \& {Steenbrugge}}]{costantini2010}
{Costantini}, E., {Kaastra}, J.~S., {Korista}, K., {et~al.} 2010, \aap, 512,
  A25

\bibitem[{{Crenshaw} {et~al.}(1995){Crenshaw}, {Boggess}, \&
  {Wu}}]{1995AJ....110.1026C}
{Crenshaw}, D.~M., {Boggess}, A., \& {Wu}, C. 1995, \aj, 110, 1026

\bibitem[{{Crenshaw} {et~al.}(1999){Crenshaw}, {Kraemer}, {Boggess}, {Maran},
  {Mushotzky}, \& {Wu}}]{crenshaw1999}
{Crenshaw}, D.~M., {Kraemer}, S.~B., {Boggess}, A., {et~al.} 1999, \apj, 516,
  750

\bibitem[{{Dadina}(2008)}]{dadina2008}
{Dadina}, M. 2008, \aap, 485, 417

\bibitem[{{Dadina} {et~al.}(2005){Dadina}, {Cappi}, {Malaguti}, {Ponti}, \& {de
  Rosa}}]{dadina2005}
{Dadina}, M., {Cappi}, M., {Malaguti}, G., {Ponti}, G., \& {de Rosa}, A. 2005,
  \aap, 442, 461

\bibitem[{{Dahari} \& {De Robertis}(1988)}]{1988ApJS...67..249D}
{Dahari}, O. \& {De Robertis}, M.~M. 1988, \apjs, 67, 249

\bibitem[{{Daly} \& {Loeb}(1990)}]{daly1990}
{Daly}, R.~A. \& {Loeb}, A. 1990, \apj, 364, 451

\bibitem[{{Danese} {et~al.}(1992){Danese}, {Zitelli}, {Granato}, {Wade}, {de
  Zotti}, \& {Mandolesi}}]{1992ApJ...399...38D}
{Danese}, L., {Zitelli}, V., {Granato}, G.~L., {et~al.} 1992, \apj, 399, 38

\bibitem[{{de Bruyn} \& {Sargent}(1978)}]{1978AJ.....83.1257D}
{de Bruyn}, A.~G. \& {Sargent}, W.~L.~W. 1978, \aj, 83, 1257

\bibitem[{{Detmers} {et~al.}(2010){Detmers}, {Kaastra}, {Costantini},
  {Verbunt}, {Cappi}, \& {de Vries}}]{detmers2010}
{Detmers}, R.~G., {Kaastra}, J.~S., {Costantini}, E., {et~al.} 2010, \aap, 516,
  A61

\bibitem[{{Detmers} {et~al.}(2011){Detmers}, {Kaastra}, {Steenbrugge},
  {Ebrero}, {Kriss}, \& {et~al.}}]{detmers2011}
{Detmers}, R.~G., {Kaastra}, J.~S., {Steenbrugge}, K.~C., {et~al.} 2011, \aap,
  in press (paper III)

\bibitem[{{Dil} {et~al.}(1981){Dil}, {Primini}, {Basinska}, {Bautz}, {Howe},
  {Lang}, {Levine}, {Lewin}, {Worrall}, {Nolan}, \&
  {Matteson}}]{1981ApJ...250..513D}
{Dil}, S., {Primini}, F.~A., {Basinska}, E., {et~al.} 1981, \apj, 250, 513

\bibitem[{{Dower} {et~al.}(1980){Dower}, {Bradt}, {Doxsey}, {Johnston}, \&
  {Griffiths}}]{1980ApJ...235..355D}
{Dower}, R.~G., {Bradt}, H.~V., {Doxsey}, R.~E., {Johnston}, M.~D., \&
  {Griffiths}, R.~E. 1980, \apj, 235, 355

\bibitem[{{Dunn} {et~al.}(2010){Dunn}, {Bautista}, {Arav}, {Moe}, {Korista},
  {Costantini}, {Benn}, {Ellison}, \& {Edmonds}}]{dunn2010}
{Dunn}, J.~P., {Bautista}, M., {Arav}, N., {et~al.} 2010, \apj, 709, 611

\bibitem[{{Ebrero} {et~al.}(2011){Ebrero}, {Kriss}, {Kaastra}, {Detmers},
  {Steenbrugge}, \& {et~al.}}]{ebrero2011}
{Ebrero}, J., {Kriss}, G.~A., {Kaastra}, J.~S., {et~al.} 2011, \aap, in press
  (paper V)

\bibitem[{{Ferland} {et~al.}(1998){Ferland}, {Korista}, {Verner}, {Ferguson},
  {Kingdon}, \& {Verner}}]{ferland1998}
{Ferland}, G.~J., {Korista}, K.~T., {Verner}, D.~A., {et~al.} 1998, \pasp, 110,
  761

\bibitem[{{Fuentes-Williams} \& {Stocke}(1988)}]{1988AJ.....96.1235F}
{Fuentes-Williams}, T. \& {Stocke}, J.~T. 1988, \aj, 96, 1235

\bibitem[{{Furlanetto} \& {Loeb}(2001)}]{furlanetto2001}
{Furlanetto}, S.~R. \& {Loeb}, A. 2001, \apj, 556, 619

\bibitem[{{Gabel} {et~al.}(2005){Gabel}, {Kraemer}, {Crenshaw}, {George},
  {Brandt}, {Hamann}, {Kaiser}, {Kaspi}, {Kriss}, {Mathur}, {Nandra}, {Netzer},
  {Peterson}, {Shields}, {Turner}, \& {Zheng}}]{gabel2005}
{Gabel}, J.~R., {Kraemer}, S.~B., {Crenshaw}, D.~M., {et~al.} 2005, \apj, 631,
  741

\bibitem[{{Glass}(2004)}]{2004MNRAS.350.1049G}
{Glass}, I.~S. 2004, \mnras, 350, 1049

\bibitem[{{Glass} {et~al.}(1982){Glass}, {Moorwood}, \&
  {Eichendorf}}]{1982A&A...107..276G}
{Glass}, I.~S., {Moorwood}, A.~F.~M., \& {Eichendorf}, W. 1982, \aap, 107, 276

\bibitem[{{Haardt}(1993)}]{haardt1993}
{Haardt}, F. 1993, \apj, 413, 680

\bibitem[{{Haardt} \& {Maraschi}(1991)}]{haardt1991}
{Haardt}, F. \& {Maraschi}, L. 1991, \apjl, 380, L51

\bibitem[{{Haardt} {et~al.}(1997){Haardt}, {Maraschi}, \&
  {Ghisellini}}]{haardt1997}
{Haardt}, F., {Maraschi}, L., \& {Ghisellini}, G. 1997, \apj, 476, 620

\bibitem[{{Hamann} {et~al.}(2001){Hamann}, {Barlow}, {Chaffee}, {Foltz}, \&
  {Weymann}}]{hamann2001}
{Hamann}, F.~W., {Barlow}, T.~A., {Chaffee}, F.~C., {Foltz}, C.~B., \&
  {Weymann}, R.~J. 2001, \apj, 550, 142

\bibitem[{{Huchra} {et~al.}(1993){Huchra}, {Latham}, {da Costa}, {Pellegrini},
  \& {Willmer}}]{huchra1993}
{Huchra}, J., {Latham}, D.~W., {da Costa}, L.~N., {Pellegrini}, P.~S., \&
  {Willmer}, C.~N.~A. 1993, \aj, 105, 1637

\bibitem[{{Jourdain} {et~al.}(1992){Jourdain}, {Bassani}, {Bouchet}, {Mandrou},
  {Ballet}, {Lebrun}, {Paul}, {Laurent}, {Churazov}, {Gilfanov}, {Sunyaev},
  {Dyackhov}, {Khavenson}, {Chulkov}, {Kuznetsov}, \& {Novikov}}]{jourdain1992}
{Jourdain}, E., {Bassani}, L., {Bouchet}, L., {et~al.} 1992, \aap, 256, L38

\bibitem[{{Kaastra} {et~al.}(2011){Kaastra}, {De Vries}, {Steenbrugge},
  {Detmers}, {Ebrero}, \& {et al.}}]{kaastra2011}
{Kaastra}, J.~S., {De Vries}, C.~P., {Steenbrugge}, K.~C., {et~al.} 2011, \aap,
  submitted (paper II)

\bibitem[{{Kaastra} {et~al.}(2004){Kaastra}, {Raassen}, {Mewe}, {Arav},
  {Behar}, {Costantini}, {Gabel}, {Kriss}, {Proga}, {Sako}, \&
  {Steenbrugge}}]{kaastra2004}
{Kaastra}, J.~S., {Raassen}, A.~J.~J., {Mewe}, R., {et~al.} 2004, \aap, 428, 57

\bibitem[{{Kaastra} {et~al.}(2002){Kaastra}, {Steenbrugge}, {Raassen}, {van der
  Meer}, {Brinkman}, {Liedahl}, {Behar}, \& {de Rosa}}]{kaastra2002}
{Kaastra}, J.~S., {Steenbrugge}, K.~C., {Raassen}, A.~J.~J., {et~al.} 2002,
  \aap, 386, 427

\bibitem[{{King}(2003)}]{king2003}
{King}, A. 2003, \apjl, 596, L27

\bibitem[{{Kinney} {et~al.}(1996){Kinney}, {Calzetti}, {Bohlin}, {McQuade},
  {Storchi-Bergmann}, \& {Schmitt}}]{kinney1996}
{Kinney}, A.~L., {Calzetti}, D., {Bohlin}, R.~C., {et~al.} 1996, \apj, 467, 38

\bibitem[{{Kirhakos} \& {Steiner}(1990)}]{1990AJ.....99.1435K}
{Kirhakos}, S.~D. \& {Steiner}, J.~E. 1990, \aj, 99, 1435

\bibitem[{{Kopylov} {et~al.}(1974){Kopylov}, {Lipovetskii}, {Pronik}, \&
  {Chuvaev}}]{1974Afz....10..483K}
{Kopylov}, I.~M., {Lipovetskii}, V.~A., {Pronik}, V.~I., \& {Chuvaev}, K.~K.
  1974, Astrofizika, 10, 483

\bibitem[{{Kotilainen} {et~al.}(1992){Kotilainen}, {Ward}, {Boisson}, {Depoy},
  \& {Smith}}]{1992MNRAS.256..149K}
{Kotilainen}, J.~K., {Ward}, M.~J., {Boisson}, C., {Depoy}, D.~L., \& {Smith},
  M.~G. 1992, \mnras, 256, 149

\bibitem[{{Kraemer} {et~al.}(2006){Kraemer}, {Crenshaw}, {Gabel}, {Kriss},
  {Netzer}, {Peterson}, {George}, {Gull}, {Hutchings}, {Mushotzky}, \&
  {Turner}}]{kraemer2006}
{Kraemer}, S.~B., {Crenshaw}, D.~M., {Gabel}, J.~R., {et~al.} 2006, \apjs, 167,
  161

\bibitem[{{Kraemer} {et~al.}(2003){Kraemer}, {Crenshaw}, {Yaqoob}, {McKernan},
  {Gabel}, {George}, {Turner}, \& {Dunn}}]{kraemer2003}
{Kraemer}, S.~B., {Crenshaw}, D.~M., {Yaqoob}, T., {et~al.} 2003, \apj, 582,
  125

\bibitem[{{Kriss} {et~al.}(2011){Kriss}, {Arav}, {Kaastra}, {Ebrero}, {Pinto},
  \& {et~al.}}]{kriss2011}
{Kriss}, G.~A., {Arav}, N., {Kaastra}, J.~S., {et~al.} 2011, \aap, in press
  (paper VI)

\bibitem[{{Kriss} {et~al.}(2000){Kriss}, {Green}, {Brotherton}, {Oegerle},
  {Sembach}, {Davidsen}, {Friedman}, {Kaiser}, {Zheng}, {Woodgate},
  {Hutchings}, {Shull}, \& {York}}]{kriss2000}
{Kriss}, G.~A., {Green}, R.~F., {Brotherton}, M., {et~al.} 2000, \apjl, 538,
  L17

\bibitem[{{Kronberg} {et~al.}(2001){Kronberg}, {Dufton}, {Li}, \&
  {Colgate}}]{kronberg2001}
{Kronberg}, P.~P., {Dufton}, Q.~W., {Li}, H., \& {Colgate}, S.~A. 2001, \apj,
  560, 178

\bibitem[{{Krongold} {et~al.}(2003){Krongold}, {Nicastro}, {Brickhouse},
  {Elvis}, {Liedahl}, \& {Mathur}}]{krongold2003}
{Krongold}, Y., {Nicastro}, F., {Brickhouse}, N.~S., {et~al.} 2003, \apj, 597,
  832

\bibitem[{{Krongold} {et~al.}(2007){Krongold}, {Nicastro}, {Elvis},
  {Brickhouse}, {Binette}, {Mathur}, \&
  {Jim{\'e}nez-Bail{\'o}n}}]{krongold2007}
{Krongold}, Y., {Nicastro}, F., {Elvis}, M., {et~al.} 2007, \apj, 659, 1022

\bibitem[{{Landt} {et~al.}(2011){Landt}, {Elvis}, {Ward}, {Bentz}, {Korista},
  \& {Karovska}}]{landt2011}
{Landt}, H., {Elvis}, M., {Ward}, M.~J., {et~al.} 2011, \mnras, in press
  (ArXiv1101.3342)

\bibitem[{{Liu}(1983)}]{1983AcApS...3..113L}
{Liu}, R. 1983, Acta Astrophysica Sinica, 3, 113

\bibitem[{{Lodders} {et~al.}(2009){Lodders}, {Palme}, \& {Gail}}]{lodders2009}
{Lodders}, K., {Palme}, H., \& {Gail}, H. 2009, in Landolt-B{\"o}rnstein -
  Group VI Astronomy and Astrophysics Numerical Data and Functional
  Relationships in Science and Technology Volume 4B: Solar System. Edited by
  J.E. Tr{\"u}mper, 2009, 4.4., 44

\bibitem[{{Longinotti} {et~al.}(2010){Longinotti}, {Costantini}, {Petrucci},
  {Boisson}, {Mouchet}, {Santos-Lleo}, {Matt}, {Ponti}, \& {Gon{\c
  c}alves}}]{longinotti2010}
{Longinotti}, A.~L., {Costantini}, E., {Petrucci}, P.~O., {et~al.} 2010, \aap,
  510, A92

\bibitem[{{MacKenty}(1990)}]{1990ApJS...72..231M}
{MacKenty}, J.~W. 1990, \apjs, 72, 231

\bibitem[{{Magnitskaia} \& {Saakian}(1976)}]{1976Afz....12..431M}
{Magnitskaia}, O.~V. \& {Saakian}, K.~A. 1976, Astrofizika, 12, 431

\bibitem[{{Maisack} {et~al.}(1993){Maisack}, {Johnson}, {Kinzer}, {Strickman},
  {Kurfess}, {Cameron}, {Jung}, {Grabelsky}, {Purcell}, \&
  {Ulmer}}]{maisack1993}
{Maisack}, M., {Johnson}, W.~N., {Kinzer}, R.~L., {et~al.} 1993, \apjl, 407,
  L61

\bibitem[{{Malzac} \& {Jourdain}(2000)}]{malzac2000}
{Malzac}, J. \& {Jourdain}, E. 2000, \aap, 359, 843

\bibitem[{{Markarian}(1973)}]{1973Afz.....9....5M}
{Markarian}, B.~E. 1973, Astrofizika, 9, 5

\bibitem[{{Markarian} \& {Lipovetskij}(1973)}]{1973Afz.....9..487M}
{Markarian}, B.~E. \& {Lipovetskij}, V.~A. 1973, Astrofizika, 9, 487

\bibitem[{{Martin} {et~al.}(1983){Martin}, {Thompson}, {Maza}, \&
  {Angel}}]{1983ApJ...266..470M}
{Martin}, P.~G., {Thompson}, I.~B., {Maza}, J., \& {Angel}, J.~R.~P. 1983,
  \apj, 266, 470

\bibitem[{{McAlary} {et~al.}(1983){McAlary}, {McLaren}, {McGonegal}, \&
  {Maza}}]{1983ApJS...52..341M}
{McAlary}, C.~W., {McLaren}, R.~A., {McGonegal}, R.~J., \& {Maza}, J. 1983,
  \apjs, 52, 341

\bibitem[{{Mediavilla} {et~al.}(1998){Mediavilla}, {Arribas}, {Garcia-Lorenzo},
  \& {del Burgo}}]{1998ApJ...494L...9M}
{Mediavilla}, E., {Arribas}, S., {Garcia-Lorenzo}, B., \& {del Burgo}, C. 1998,
  \apjl, 494, L9

\bibitem[{{Mehdipour} {et~al.}(2011){Mehdipour}, {Branduardi-Raymont},
  {Kaastra}, {Petrucci}, {Kriss}, \& {et~al.}}]{mehdipour2011}
{Mehdipour}, M., {Branduardi-Raymont}, G., {Kaastra}, J.~S., {et~al.} 2011,
  \aap, in press (paper IV)

\bibitem[{{Mingaliev} {et~al.}(1978){Mingaliev}, {Pustilnik}, {Trushkin},
  {Kirakosian}, \& {Malumian}}]{1978Afz....14...91M}
{Mingaliev}, M.~G., {Pustilnik}, S.~A., {Trushkin}, S.~A., {Kirakosian}, R.~M.,
  \& {Malumian}, V.~G. 1978, Astrofizika, 14, 91

\bibitem[{{Moe} {et~al.}(2009){Moe}, {Arav}, {Bautista}, \&
  {Korista}}]{moe2009}
{Moe}, M., {Arav}, N., {Bautista}, M.~A., \& {Korista}, K.~T. 2009, \apj, 706,
  525

\bibitem[{{Moorwood}(1986)}]{1986A&A...166....4M}
{Moorwood}, A.~F.~M. 1986, \aap, 166, 4

\bibitem[{{Morini} {et~al.}(1987){Morini}, {Lipani}, \&
  {Molteni}}]{1987ApJ...317..145M}
{Morini}, M., {Lipani}, N.~A., \& {Molteni}, D. 1987, \apj, 317, 145

\bibitem[{{Moshir} {et~al.}(1990){Moshir}, {Kopan}, {Conrow}, {McCallon},
  {Hacking}, {Gregorich}, {Rohrbach}, {Melnyk}, {Rice}, {Fullmer}, \& {et
  al.}}]{moshir1990}
{Moshir}, M., {Kopan}, G., {Conrow}, T., {et~al.} 1990, in IRAS Faint Source
  Catalogue, version 2.0 (1990)

\bibitem[{{Murphy} {et~al.}(1995){Murphy}, {Lockman}, \& {Savage}}]{murphy1995}
{Murphy}, E.~M., {Lockman}, F.~J., \& {Savage}, B.~D. 1995, \apj, 447, 642

\bibitem[{{Mushotzky} {et~al.}(1980){Mushotzky}, {Marshall}, {Boldt}, {Holt},
  \& {Serlemitsos}}]{1980ApJ...235..377M}
{Mushotzky}, R.~F., {Marshall}, F.~E., {Boldt}, E.~A., {Holt}, S.~S., \&
  {Serlemitsos}, P.~J. 1980, \apj, 235, 377

\bibitem[{{Nandra} {et~al.}(2000){Nandra}, {Le}, {George}, {Edelson},
  {Mushotzky}, {Peterson}, \& {Turner}}]{nandra2000}
{Nandra}, K., {Le}, T., {George}, I.~M., {et~al.} 2000, \apj, 544, 734

\bibitem[{{Netzer} {et~al.}(2003){Netzer}, {Kaspi}, {Behar}, {Brandt},
  {Chelouche}, {George}, {Crenshaw}, {Gabel}, {Hamann}, {Kraemer}, {Kriss},
  {Nandra}, {Peterson}, {Shields}, \& {Turner}}]{netzer2003}
{Netzer}, H., {Kaspi}, S., {Behar}, E., {et~al.} 2003, \apj, 599, 933

\bibitem[{{O'Donnell}(1994)}]{odonnell1994}
{O'Donnell}, J.~E. 1994, \apj, 422, 158

\bibitem[{{Ogle} {et~al.}(2004){Ogle}, {Mason}, {Page}, {Salvi}, {Cordova},
  {McHardy}, \& {Priedhorsky}}]{ogle2004}
{Ogle}, P.~M., {Mason}, K.~O., {Page}, M.~J., {et~al.} 2004, \apj, 606, 151

\bibitem[{{Osterbrock}(1977)}]{1977ApJ...215..733O}
{Osterbrock}, D.~E. 1977, \apj, 215, 733

\bibitem[{{Page} {et~al.}(2003){Page}, {Davis}, \& {Salvi}}]{page2003}
{Page}, M.~J., {Davis}, S.~W., \& {Salvi}, N.~J. 2003, \mnras, 343, 1241

\bibitem[{{Perola} {et~al.}(2002){Perola}, {Matt}, {Cappi}, {Fiore},
  {Guainazzi}, {Maraschi}, {Petrucci}, \& {Piro}}]{perola2002}
{Perola}, G.~C., {Matt}, G., {Cappi}, M., {et~al.} 2002, \aap, 389, 802

\bibitem[{{Peterson} {et~al.}(1984){Peterson}, {Crenshaw}, {Meyers}, {Byard},
  \& {Foltz}}]{1984ApJ...279..529P}
{Peterson}, B.~M., {Crenshaw}, D.~M., {Meyers}, K.~A., {Byard}, P.~L., \&
  {Foltz}, C.~B. 1984, \apj, 279, 529

\bibitem[{{Peterson} {et~al.}(2004){Peterson}, {Ferrarese}, {Gilbert}, {Kaspi},
  {Malkan}, {Maoz}, {Merritt}, {Netzer}, {Onken}, {Pogge}, {Vestergaard}, \&
  {Wandel}}]{2004ApJ...613..682P}
{Peterson}, B.~M., {Ferrarese}, L., {Gilbert}, K.~M., {et~al.} 2004, \apj, 613,
  682

\bibitem[{{Peterson} {et~al.}(1982){Peterson}, {Foltz}, {Byard}, \&
  {Wagner}}]{1982ApJS...49..469P}
{Peterson}, B.~M., {Foltz}, C.~B., {Byard}, P.~L., \& {Wagner}, R.~M. 1982,
  \apjs, 49, 469

\bibitem[{{Peterson} {et~al.}(1998){Peterson}, {Wanders}, {Bertram}, {Hunley},
  {Pogge}, \& {Wagner}}]{1998ApJ...501...82P}
{Peterson}, B.~M., {Wanders}, I., {Bertram}, R., {et~al.} 1998, \apj, 501, 82

\bibitem[{{Petrucci} {et~al.}(2001){Petrucci}, {Haardt}, {Maraschi}, {Grandi},
  {Malzac}, {Matt}, {Nicastro}, {Piro}, {Perola}, \& {De Rosa}}]{petrucci2001}
{Petrucci}, P.~O., {Haardt}, F., {Maraschi}, L., {et~al.} 2001, \apj, 556, 716

\bibitem[{{Petrucci} {et~al.}(2000){Petrucci}, {Haardt}, {Maraschi}, {Grandi},
  {Matt}, {Nicastro}, {Piro}, {Perola}, \& {De Rosa}}]{petrucci2000}
{Petrucci}, P.~O., {Haardt}, F., {Maraschi}, L., {et~al.} 2000, \apj, 540, 131

\bibitem[{{Petrucci} {et~al.}(2004){Petrucci}, {Maraschi}, {Haardt}, \&
  {Nandra}}]{petrucci2004}
{Petrucci}, P.~O., {Maraschi}, L., {Haardt}, F., \& {Nandra}, K. 2004, \aap,
  413, 477

\bibitem[{{Petrucci} {et~al.}(2011)}]{petrucci2011}
{Petrucci}, P.-O. {et~al.} 2011, \aap, in preparation

\bibitem[{{Phillips} {et~al.}(1983){Phillips}, {Baldwin}, {Atwood}, \&
  {Carswell}}]{1983ApJ...274..558P}
{Phillips}, M.~M., {Baldwin}, J.~A., {Atwood}, B., \& {Carswell}, R.~F. 1983,
  \apj, 274, 558

\bibitem[{{Pietsch} {et~al.}(1981){Pietsch}, {Reppin}, {Truemper}, {Voges},
  {Lewin}, {Kendziorra}, \& {Staubert}}]{1981A&A....94..234P}
{Pietsch}, W., {Reppin}, C., {Truemper}, J., {et~al.} 1981, \aap, 94, 234

\bibitem[{{Platania} {et~al.}(2002){Platania}, {Burigana}, {De Zotti},
  {Lazzaro}, \& {Bersanelli}}]{platania2002}
{Platania}, P., {Burigana}, C., {De Zotti}, G., {Lazzaro}, E., \& {Bersanelli},
  M. 2002, \mnras, 337, 242

\bibitem[{{Ponti} {et~al.}(2009){Ponti}, {Cappi}, {Vignali}, {Miniutti},
  {Tombesi}, {Dadina}, {Fabian}, {Grandi}, {Kaastra}, {Petrucci}, {Bianchi},
  {Matt}, {Maraschi}, \& {Malaguti}}]{ponti2009}
{Ponti}, G., {Cappi}, M., {Vignali}, C., {et~al.} 2009, \mnras, 394, 1487

\bibitem[{{Pounds} {et~al.}(2001){Pounds}, {Reeves}, {O'Brien}, {Page},
  {Turner}, \& {Nayakshin}}]{pounds2001}
{Pounds}, K., {Reeves}, J., {O'Brien}, P., {et~al.} 2001, \apj, 559, 181

\bibitem[{{Pounds} {et~al.}(1994){Pounds}, {Nandra}, {Fink}, \&
  {Makino}}]{1994MNRAS.267..193P}
{Pounds}, K.~A., {Nandra}, K., {Fink}, H.~H., \& {Makino}, F. 1994, \mnras,
  267, 193

\bibitem[{{Poutanen} \& {Svensson}(1996)}]{poutanen1996}
{Poutanen}, J. \& {Svensson}, R. 1996, \apj, 470, 249

\bibitem[{{Reeves} {et~al.}(2004){Reeves}, {Nandra}, {George}, {Pounds},
  {Turner}, \& {Yaqoob}}]{reeves2004}
{Reeves}, J.~N., {Nandra}, K., {George}, I.~M., {et~al.} 2004, \apj, 602, 648

\bibitem[{{Rieke}(1978)}]{1978ApJ...226..550R}
{Rieke}, G.~H. 1978, \apj, 226, 550

\bibitem[{{Roche} {et~al.}(1984){Roche}, {Whitmore}, {Aitken}, \&
  {Phillips}}]{1984MNRAS.207...35R}
{Roche}, P.~F., {Whitmore}, B., {Aitken}, D.~K., \& {Phillips}, M.~M. 1984,
  \mnras, 207, 35

\bibitem[{{Rosenblatt} {et~al.}(1992){Rosenblatt}, {Malkan}, {Sargent}, \&
  {Readhead}}]{1992ApJS...81...59R}
{Rosenblatt}, E.~I., {Malkan}, M.~A., {Sargent}, W.~L.~W., \& {Readhead},
  A.~C.~S. 1992, \apjs, 81, 59

\bibitem[{{Rothschild} {et~al.}(1983){Rothschild}, {Baity}, {Gruber},
  {Matteson}, {Peterson}, \& {Mushotzky}}]{1983ApJ...269..423R}
{Rothschild}, R.~E., {Baity}, W.~A., {Gruber}, D.~E., {et~al.} 1983, \apj, 269,
  423

\bibitem[{{Scannapieco} \& {Oh}(2004)}]{scannapieco2004}
{Scannapieco}, E. \& {Oh}, S.~P. 2004, \apj, 608, 62

\bibitem[{{Shang} {et~al.}(2005){Shang}, {Brotherton}, {Green}, {Kriss},
  {Scott}, {Quijano}, {Blaes}, {Hubeny}, {Hutchings}, {Kaiser}, {Koratkar},
  {Oegerle}, \& {Zheng}}]{shang2005}
{Shang}, Z., {Brotherton}, M.~S., {Green}, R.~F., {et~al.} 2005, \apj, 619, 41

\bibitem[{{Silk} \& {Rees}(1998)}]{silk1998}
{Silk}, J. \& {Rees}, M.~J. 1998, \aap, 331, L1

\bibitem[{{Singh} {et~al.}(1985){Singh}, {Garmire}, \&
  {Nousek}}]{1985ApJ...297..633S}
{Singh}, K.~P., {Garmire}, G.~P., \& {Nousek}, J. 1985, \apj, 297, 633

\bibitem[{{Singh} \& {Westergaard}(1992)}]{1992A&A...264..489S}
{Singh}, K.~P. \& {Westergaard}, N.~J. 1992, \aap, 264, 489

\bibitem[{{Smith} {et~al.}(2007){Smith}, {Page}, \&
  {Branduardi-Raymont}}]{smith2007}
{Smith}, R.~A.~N., {Page}, M.~J., \& {Branduardi-Raymont}, G. 2007, \aap, 461,
  135

\bibitem[{{Steenbrugge} {et~al.}(2009){Steenbrugge}, {Fenov{\v c}{\'{\i}}k},
  {Kaastra}, {Costantini}, \& {Verbunt}}]{steenbrugge2009}
{Steenbrugge}, K.~C., {Fenov{\v c}{\'{\i}}k}, M., {Kaastra}, J.~S.,
  {Costantini}, E., \& {Verbunt}, F. 2009, \aap, 496, 107

\bibitem[{{Steenbrugge} {et~al.}(2005){Steenbrugge}, {Kaastra}, {Crenshaw},
  {Kraemer}, {Arav}, {George}, {Liedahl}, {van der Meer}, {Paerels}, {Turner},
  \& {Yaqoob}}]{steenbrugge2005}
{Steenbrugge}, K.~C., {Kaastra}, J.~S., {Crenshaw}, D.~M., {et~al.} 2005, \aap,
  434, 569

\bibitem[{{Steenbrugge} {et~al.}(2003){Steenbrugge}, {Kaastra}, {de Vries}, \&
  {Edelson}}]{steenbrugge2003}
{Steenbrugge}, K.~C., {Kaastra}, J.~S., {de Vries}, C.~P., \& {Edelson}, R.
  2003, \aap, 402, 477

\bibitem[{{Stein} \& {Weedman}(1976)}]{1976ApJ...205...44S}
{Stein}, W.~A. \& {Weedman}, D.~W. 1976, \apj, 205, 44

\bibitem[{{Stern} {et~al.}(1995){Stern}, {Poutanen}, {Svensson}, {Sikora}, \&
  {Begelman}}]{stern1995}
{Stern}, B.~E., {Poutanen}, J., {Svensson}, R., {Sikora}, M., \& {Begelman},
  M.~C. 1995, \apjl, 449, L13

\bibitem[{{Svensson}(1996)}]{svensson1996}
{Svensson}, R. 1996, \aaps, 120, C475

\bibitem[{{Ulvestad} \& {Wilson}(1984)}]{1984ApJ...278..544U}
{Ulvestad}, J.~S. \& {Wilson}, A.~S. 1984, \apj, 278, 544

\bibitem[{{Unger} {et~al.}(1987){Unger}, {Lawrence}, {Wilson}, {Elvis}, \&
  {Wright}}]{1987MNRAS.228..521U}
{Unger}, S.~W., {Lawrence}, A., {Wilson}, A.~S., {Elvis}, M., \& {Wright},
  A.~E. 1987, \mnras, 228, 521

\bibitem[{{Winge} {et~al.}(2000){Winge}, {Storchi-Bergmann}, {Ward}, \&
  {Wilson}}]{2000MNRAS.316....1W}
{Winge}, C., {Storchi-Bergmann}, T., {Ward}, M.~J., \& {Wilson}, A.~S. 2000,
  \mnras, 316, 1

\bibitem[{{Wu} {et~al.}(1980){Wu}, {Boggess}, \& {Gull}}]{1980ApJ...242...14W}
{Wu}, C., {Boggess}, A., \& {Gull}, T.~R. 1980, \apj, 242, 14

\bibitem[{{Wu} {et~al.}(2000){Wu}, {Fabian}, \& {Nulsen}}]{wu2000}
{Wu}, K.~K.~S., {Fabian}, A.~C., \& {Nulsen}, P.~E.~J. 2000, \mnras, 318, 889

\bibitem[{{Wu} {et~al.}(2009){Wu}, {Charmandaris}, {Huang}, {Spinoglio}, \&
  {Tommasin}}]{wu2009}
{Wu}, Y., {Charmandaris}, V., {Huang}, J., {Spinoglio}, L., \& {Tommasin}, S.
  2009, \apj, 701, 658

\bibitem[{{Wyithe} \& {Loeb}(2003)}]{wyithe2003}
{Wyithe}, J.~S.~B. \& {Loeb}, A. 2003, \apj, 595, 614

\bibitem[{{Zdziarski} {et~al.}(1990){Zdziarski}, {Ghisellini}, {George},
  {Fabian}, {Svensson}, \& {Done}}]{zdziarski1990}
{Zdziarski}, A.~A., {Ghisellini}, G., {George}, I.~M., {et~al.} 1990, \apjl,
  363, L1

\bibitem[{{Zdziarski} \& {Grandi}(2001)}]{zdziarski2001}
{Zdziarski}, A.~A. \& {Grandi}, P. 2001, \apj, 551, 186

\end{thebibliography}

\end{document}